\documentclass[preprint,12pt]{elsarticle}
\journal{journal} \RequirePackage{graphicx}

\begin{document}

\begin{frontmatter}

\title{Compact stars: a generalized model}

\author{S.K. Maurya}
\address{Department of Mathematical and Physical Sciences,
College of Arts and Science, University of Nizwa, Nizwa, Sultanate
of Oman\\sunil@unizwa.edu.om}

\author{Debabrata Deb}
\address{Department of
Physics, Indian Institute of Engineering Science \& Technology, Shibpur, Howrah 711103, West Bengal, India\\d.deb32@gmail.com}

\author{Saibal Ray}
\address{Department of Physics,
Government College of Engineering \& Ceramic Technology, Kolkata 700010, West Bengal, India
\\saibal@associates.iucaa.in}

\author{P.K.F. Kuhfittig}
\address{Department of Mathematics, 
Milwaukee School of Engineering, Milwaukee, Wisconsin 53202-3109, USA
\\kuhfitti@msoe.edu}

\date{Received: date / Accepted: date}

\begin{abstract}
This paper discusses a generalized model for compact stars, assumed to be anisotropic in nature due to the
spherical symmetry and high density.  After embedding the four-dimensional spacetime in a five-dimensional
flat spacetime, which may be treated as an alternative to Karmarkar's condition of embedding class 1 spacetime, 
the Einstein field equations were solved by employing a class of physically acceptable
metric functions proposed by Lake \cite{Lake2003}.  The physical properties determined include the anisotropic factor
showing that the anisotropy is zero at the center and maximal at the surface.  Other boundary conditions
yielded the values of various parameters needed for rendering the numerous plots and also led to the
EOS parameters.  It was further determined that the usual energy conditions are satisfied and that the
compact structures are stable, based on several criteria, starting with the TOV equation.  The calculation
of the effective gravitational mass shows that the models satisfy the Buchdahl condition.  Finally, the
values of the numerous constants and physical parameters were determined specifically for the strange star LMCX-4. 
It is shown that the present generalized model can justify most off the compact stars including white dwarfs and 
ultra dense compact stars for a suitable tuning of the parametric values of $n$.
\end{abstract}


\end{frontmatter}

\section{Introduction}\label{sec1}
In the course of many years embedding theorems have remained a topic of interest among theoretical physicists and mathematicians. Using this theory one may connect the classical general theory of relativity to the higher-dimensional manifolds and thereby explain the internal symmetry groups characteristic of particles. Using Campbell's theorem~\cite{Campbell1926}, Tavakol and his group~\cite{Tavakol1,Tavakol2,Tavakol3} have connected different manifolds having a difference of one dimension with a ladder, and in the lower end the $4D$ Einstein equations are embedded in a 5D Ricci-flat spacetime. This provides the algebraic justifications for the induced-matter theory and membrane theory~\cite{Wesson2006,Wesson1992}. It is worth mentioning that $5$ dimensions for particles play a role as the low energy limit of certain higher-dimensional theories such as $10$D super symmetry, $11$D supper gravity and higher D-dimensional string theory and the idea is widely accepted as the simple extension of spacetime, not space-time or space time.

If $n+m$ is the lowest number of the flat space in which $n$-dimensional Riemannian space is embedded, then the latter is referred to as class $m$. It is well known that the interior Schwarzschild solution and the Friedmann universes are of class one, whereas the exterior Schwarzschild solution is a Riemannian metric of class two. Collinson~\cite{Collinson1968,Collinson1966} showed $6$-dimensional embeddings of the plane-fronted waves and proposed a vacuum spacetime of class two. Whereas Szekeres~\cite{Szekeres1966} denied the presence of the vacuum spacetime of class one, he showed that the de Sitter cosmological models are the only class one Einstein spaces. He also showed that for the Friedmann cosmologies, the perfect-fluid flows of class one have vanishing pressure and are hyper surface orthogonal. When the gravitational and electromagnetic fields are null, Einstein-Maxwell fields are of class one.

In the present article we have assumed a general spherically symmetric metric of class two and with a suitable choice of a coordinate system we have reduced this metric to class one. Using this spherically symmetric metric of class one, we have produced a generalized model for any compact star by applying the formalism previously presented by Maurya and his collaborators~\cite{Maurya1,Maurya2,Maurya3,Maurya4,Maurya5,Maurya6,Maurya7,Maurya8}. As a special note we like to mention that for the study of different compact stars we have not considered any specific equation of state (EOS) parameter. We find that EOS parameters vary from case to case for the different compact stars.

Following the works of many scientists~\cite{Bowers1974,Ruderman1972,Schunck2003,Herrera1997,Herrera2004,Herrera2008,Ivanov2002} one may obtain some detailed ideas on the anisotropic features of the highly dense spherically symmetric stellar models. In this article we assume our system to be a highly dense spherically symmetric fluid sphere and that the pressure is anisotropic in nature. In support of our assumption we want to mention the proposal of Ruderman~\cite{Ruderman1972} who predicted that some types of nuclear matter having a high density $(> {{10}^{14}} gm/{{cm}^{3}})$, exhibit an anisotropic nature and interact in the relativistic realm. The basic reasons for arising anisotropy inside compact stars may be the presence of superfluids, the existence of mixtures of different types of fluids, rotation, the presence of any form of external or magnetic fields and phase transition, etc.

This paper has been organized as following: in Sec.~\ref{sec2} the procedure for the reduction of the general spherically symmetric metric of class two into class one is shown. In Sec.~\ref{sec3} and Sec.~\ref{sec4} the basic Einstein field equations and their solutions are shown, respectively. The boundary conditions and different physical properties of the stellar model are studied in Sec.~\ref{sec5} and Sect~\ref{sec6}. The major results of this paper are summarized in the concluding Sec.~\ref{sec7}.

\section{Spherical symmetric metric and class one condition for the metric}\label{sec2}
To describe the interior of a static and spherically symmetric object, the line element in Schwarzschild coordinates $(x^{a})=(t,r,\theta,\phi)$ can be written as (in natural units, $G=c=1$)
\begin{equation}
ds^{2}=e^{\nu(r)}dt^{2}-e^{\lambda(r)}dr^{2}-r^{2}\left(d\theta^{2}+\sin^{2}\theta d\phi^{2} \right),\label{eq1}
\end{equation}
where $\lambda$ and $\nu$ are functions of the radial coordinate $r$.

We now show that the above metric (\ref{eq1}) of class two can be reduced to class one and then embedded in the 5-dimensional flat 
metric
\begin{eqnarray}
&\qquad\qquad\hspace{-2.5cm} {ds^{2}}=-\left(dz^1\right)^2-\left(dz^2\right)^2-\left(dz^3\right)^2-\left(dz^4\right)^2+\left(dz^5\right)^2.\label{eq2}
\end{eqnarray}

We start with the following:
 $z^1=r\,sin\theta\,cos\phi$, $z^2=r\,sin\theta\,sin\phi$, $z^3=r\,cos\theta$, $z^4=\sqrt{K}\,e^{\frac{\nu}{2}}\,cosh{\frac{t}{\sqrt{K}}}$ and $z^5=\sqrt{K}\,e^{\frac{\nu}{2}}\,sinh{\frac{t}{\sqrt{K}}}$.

Then the differentials of the above components are
 \begin{eqnarray}\label{eq2a}
&\qquad \hspace{-2cm}dz^1=dr\,sin\theta\,cos\phi + r\,cos\theta\,cos\phi\,d\theta\,-r\,sin\theta\,sin\phi\,d\phi,\\ \label{eq2b}
&\qquad \hspace{-2cm}dz^2=dr\,sin\theta\,sin\phi + r\,cos\theta\,sin\phi\,d\theta\,+r\,sin\theta\,cos\phi\,d\phi,\\ \label{eq2c}
&\qquad \hspace{-2cm}dz^3=dr\,cos\theta\, - r\,sin\theta\,d\theta,\\ \label{eq2d}
&\qquad \hspace{-2cm}dz^4=\sqrt{K}\,e^{\frac{\nu}{2}}\,\frac{\nu'}{2}\,cosh{\frac{t}{\sqrt{K}}}\,dr + e^{\frac{\nu}{2}}\,sinh{\frac{t}{\sqrt{K}}}\,dt,\\ \label{eq2e}
&\qquad \hspace{-2cm}dz^5=\sqrt{K}\,e^{\frac{\nu}{2}}\,\frac{\nu'}{2}\,sinh{\frac{t}{\sqrt{K}}}\,dr + e^{\frac{\nu}{2}}\,cosh{\frac{t}{\sqrt{K}}}\,dt,   
\end{eqnarray}
where the prime denotes differentiation with respect to the radial coordinate $r$.

Then the expressions for [$-\left(dz^1\right)^2-\left(dz^2\right)^2-\left(dz^3\right)^2$] and [$-\left(dz^4\right)^2+\left(dz^5\right)^2$] can be obtained as
\begin{equation}
-\left(dz^1\right)^2-\left(dz^2\right)^2-\left(dz^3\right)^2= -dr^2-r^{2}\left(d\theta^{2}+\sin^{2}\theta d\phi^{2} \right),\label{eq3}
\end{equation}

\begin{eqnarray}
&\qquad\hspace{-1cm}-\left(dz^4\right)^2+\left(dz^5\right)^2= e^{\nu(r)}dt^{2}-\frac{K\,e^{\nu}}{4}\,{\nu'}^2\,dr^2. \label{eq4}
\end{eqnarray}

On inserting the Eq. (\ref{eq3}) and Eq. (\ref{eq4}) into the metric (\ref{eq2}), we get
\begin{eqnarray}
&\qquad\hspace{-1.5cm} ds^{2}=e^{\nu(r)}dt^{2}-\left(\,1+\frac{K\,e^{\nu}}{4}\,{\nu'}^2\,\right)\,dr^{2}-r^{2}\left(d\theta^{2}+\sin^{2}\theta d\phi^{2} \right).\nonumber \\ \label{eq5}
\end{eqnarray}

So metric (\ref{eq5}) represents metric (\ref{eq1}) if
\begin{eqnarray}
&\qquad\hspace{1cm}e^{\lambda}=\left(\,1+\frac{K\,e^{\nu}}{4}\,{\nu'}^2\,\right).\label{eq6}
\end{eqnarray}

Thus we have embedded the 4-dimensional spacetime in the flat metric of a 5-dimensional spacetime, thereby reducing the class two metric to class one.

\section{Basic field equations}\label{sec3}
To describe a spherically symmetric anisotropic fluid system, we use the general energy-momentum tensor
\begin{equation}
T_{\nu}^{\mu}=(\rho+p_t)u^{\mu}u_{\nu}-p_t g_{\nu}^{\mu}+(p_r-p_t)\eta^{\mu}\eta_{\nu},\label{eq7}
\end{equation}
with $ u^{\mu}u_{\nu} =-\eta^{\mu}\eta_{\nu} = 1 $ , where the vector $u_{\mu}$ is the fluid 4-velocity of the local rest frame and $\eta^{\mu}$ is the unit space-like vector; these are orthogonal to each other, i.e., $u^{\mu}\eta_{\nu}= 0$. Here $\rho$, $p_r$ and $p_t$  are the matter density, radial and tangential pressure, respectively, corresponding to the anisotropic fluid.

The Einstein field equations for the metric (\ref{eq1}) can be written as
\begin{eqnarray}\label{eq8a}
&\qquad \hspace{-1cm}\frac{1-e^{-\lambda}}{r^{2}}+\frac{e^{-\lambda}\lambda'}{r}=8\pi\rho,\\	\label{eq8b}
&\qquad \hspace{-1cm}\frac{e^{-\lambda}-1}{r^{2}}+\frac{e^{-\lambda}\nu'}{r}=8\pi p_{r},\\ \label{eq8c}
&\qquad \hspace{-1cm} e^{-\lambda}\left(\frac{\nu''}{2}+\frac{\nu'^{2}}{4}-\frac{\nu'\lambda'}{4}+\frac{\nu'-\lambda'}{2r} \right)=8\pi p_t.
\end{eqnarray}

\section{General solutions of the class one metric}\label{sec4}
To solve the Einstein field equations of the proposed class one metric~(\ref{eq5}), we have chosen the following monotone increasing function of $r$ that is free from a singularity at the center:
\begin{equation}
e^{\nu}=B\,(1-Ar^2)^n,\label{eq9}
\end{equation}
where $A$ and $B$ are constants and $n \le -3$. This choice is based on the discussion by Lake~\cite{Lake2003}, who showed that a single
monotone function can generate all regular spherically symmetric perfect-fluid solutions of the Einstein field equations.  Using a form that
satisfies the criteria in Ref. \cite{Lake2003} ensures that our choice of $e^{\nu}$ is physically relevant.

The reasons for the choice of $n \le -3$ are the following:\\
i) for $n=0$ the space time becomes flat;\\
ii) for $n=-1$ and -2 we are not getting a physically valid solution and causality condition is also not maintained in this case.

One may find as $n\rightarrow{-\infty}$, $e^{\nu}$ takes the form $$\lim _{n\to{-\infty} }{e^{\nu}}=B{ e^{C{r}^{2}}},$$
where $C=-nA$ (constant).

Now substituting the expression for $e^{\nu}$ from Eq. (\ref{eq9}) into Eq. (\ref{eq6}), we get
\begin{equation}
e^{\lambda}=[1+D\,Ar^2\,(1-Ar^2)^{n-2}],\label{eq10}
\end{equation}
where we suppose that $D=n^2ABK $.

\begin{figure}[h!]
\centering
    \includegraphics[width=6.7cm]{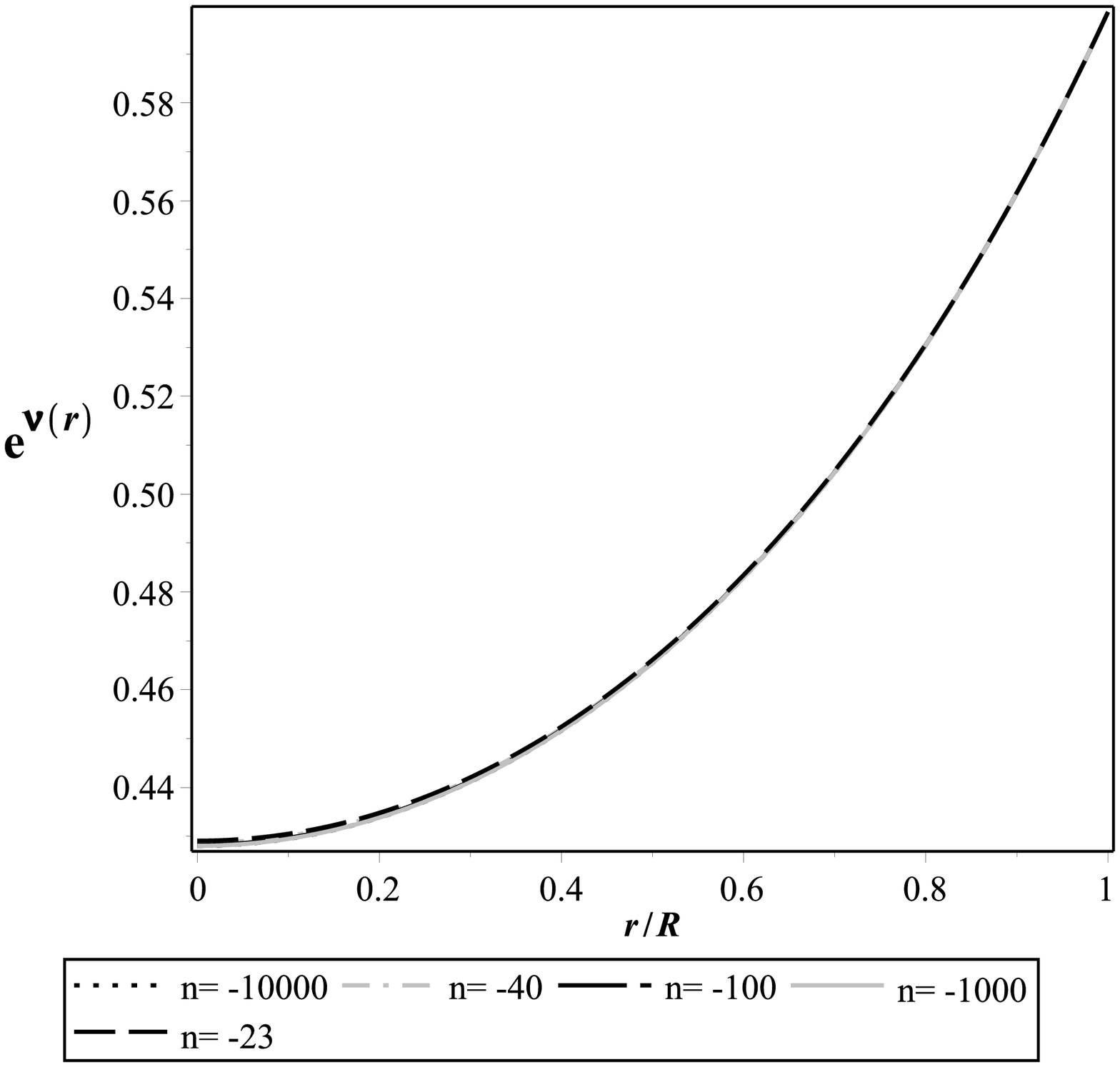}
    \includegraphics[width=6.7cm]{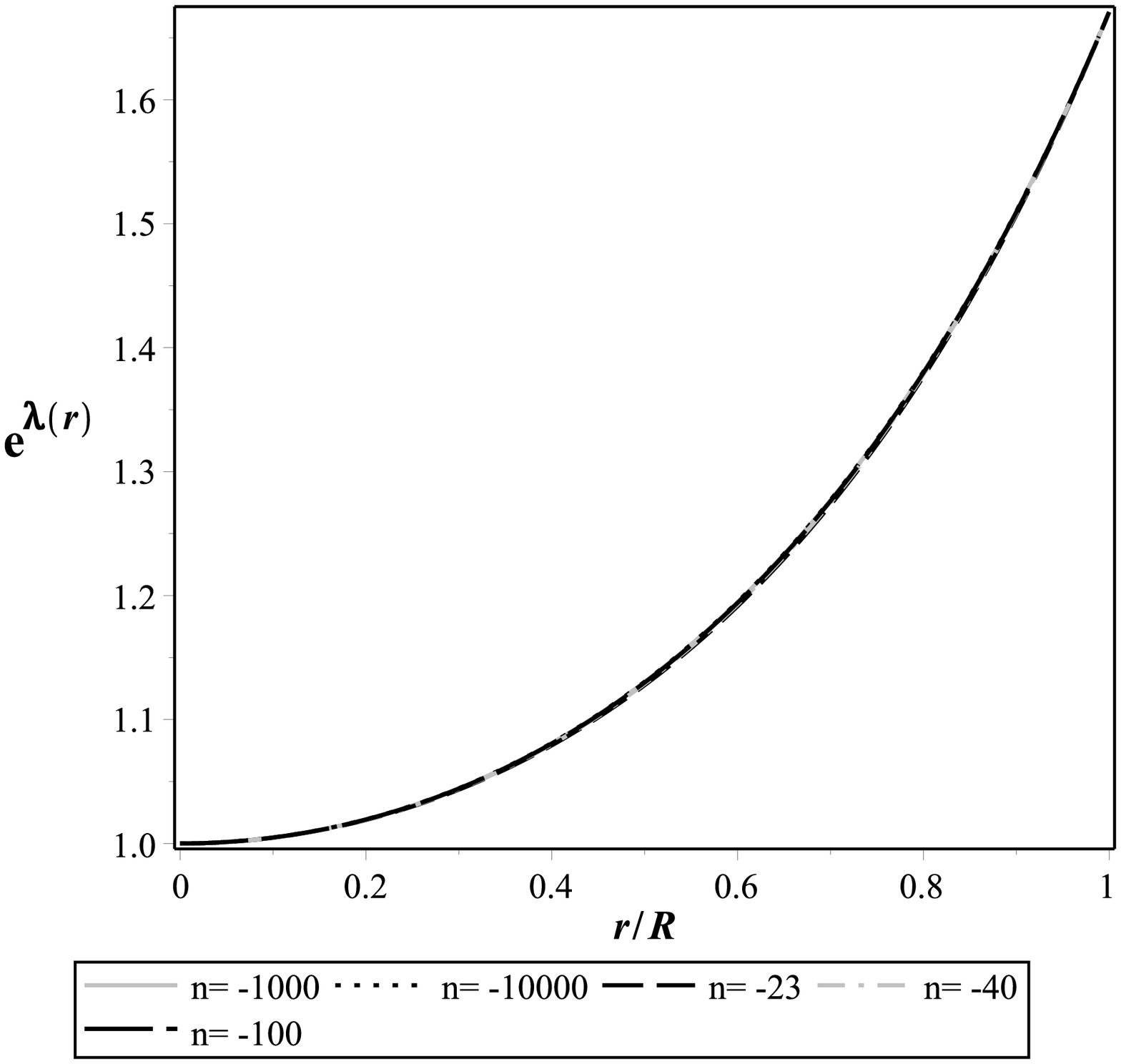}
\caption{Variation of the metric functions $e^{\nu}$ (left panel) and $e^{\lambda}$ (right panel) with the fractional coordinate $r/R$ for $LMC\,X-4$. For the purpose of plotting the graph, the values of the constants or parameters for different values of $n$ are taken from Tables~1-3 which are provided later part of the article and also will hereafter be used for all the other figures.} \label{Fig1}
\end{figure}

The expressions for the energy density, radial pressure and tangential pressure can be determined by inserting the expressions for $e^{\nu}$ and $e^{\lambda}$ from Eqs. (\ref{eq9}) and (\ref{eq10}) into Eqs.~(\ref{eq8a}-\ref{eq8c}) yielding
\begin{eqnarray}\label{eq13}
&\qquad \hspace{-2.5cm} \rho= {\frac { A\,D\, \left( 1-A{r}^{2} \right) ^{n}
 \left[ 3-2\, \left( n+1 \right) A{r}^{2}+ A\,D\,{r}^{2}
 \left( 1-A{r}^{2} \right) ^{n}+ \left( 2\,n-1 \right) {A}^{2}{r}^{4}
 \right] }{8\,\pi \, \left[  \left( 1-A{r}^{2} \right) ^{2}+ A\,D\,{r}^{2} \left( 1-A{r}^{2} \right) ^{n} \right] ^{2}}},\\\label{eq11}
&\qquad\hspace{-2.5cm} {p_r}={\frac {A \left[ 2\,n \left( -1+A{r}^{2} \right) - D\,
 \left( 1-A{r}^{2} \right) ^{n} \right] }{8\pi \, \left[
 \left( 1-A{r}^{2} \right) ^{2}+ A\,D\,{r}^{2} \left( 1-A{r}^{2} \right) ^{n} \right] }},\\ \label{eq12} 
&\qquad\hspace{-2.5cm} {p_t}={\frac {A \left( -1+A{r}^{2} \right) \left[  D\, \left( 1+A{r}^{2} \right)
 \left( 1-A{r}^{2} \right) ^{n}+n \left( -1+A{r}^{2} \right)  \left( n
A{r}^{2}-2 \right)  \right] }{8\,\pi \, \left[ A\,D\,{r}^{2} \left( 1-A{r}^{2} \right) ^{n}+
 \left( -1+A{r}^{2} \right) ^{2} \right] ^{2}}}.
 \end{eqnarray}

\begin{figure}[h!]
\centering
    \includegraphics[width=6.7cm]{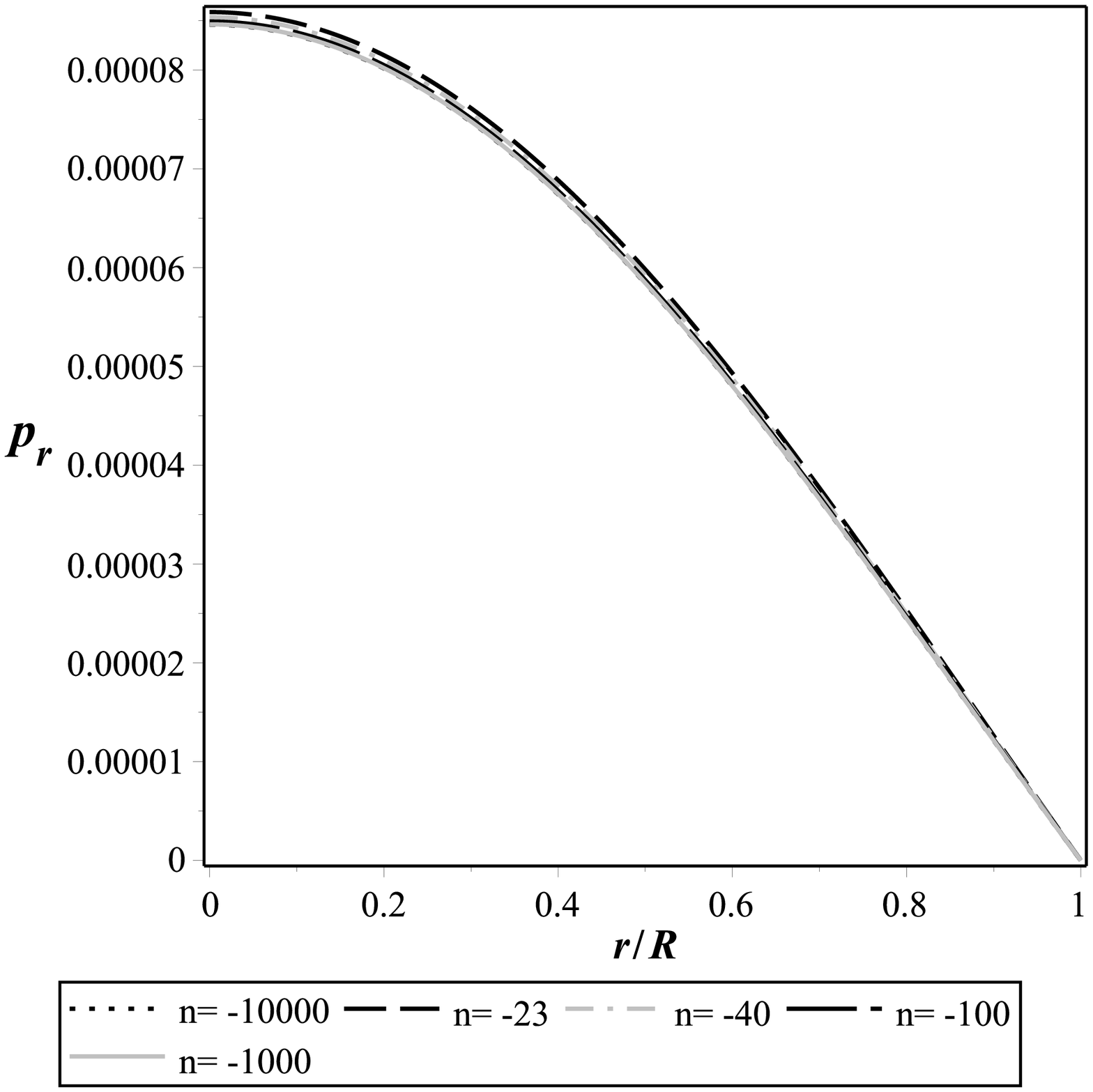}
    \includegraphics[width=6.7cm]{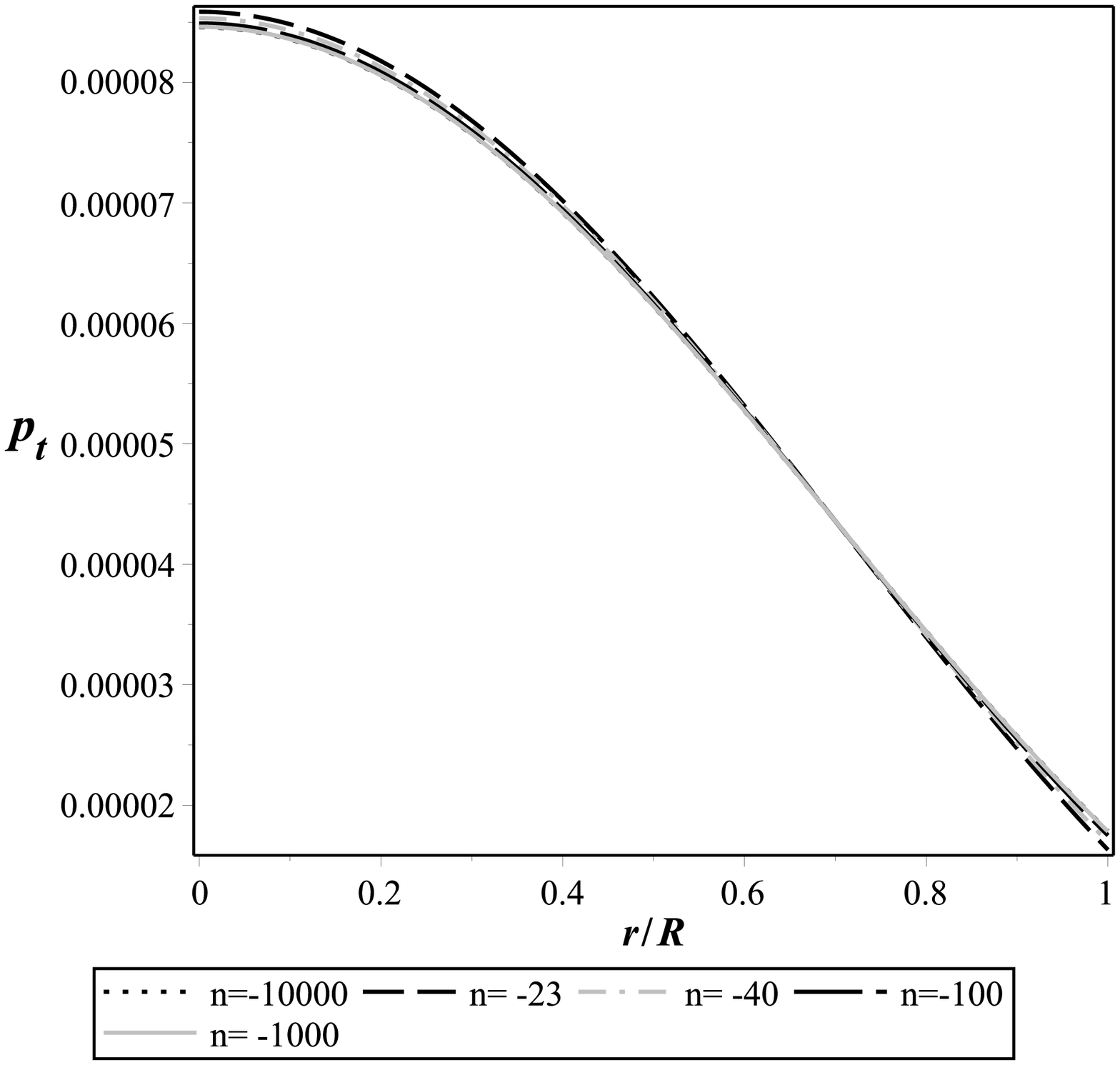}
\caption{Variation of the radial pressure, $p_r$, (left panel) and the tangential pressure, $p_t$, (right panel) with the fractional coordinate $r/R$ for $LMC\,X-4$. }
    \label{Fig2}
\end{figure}

\begin{figure}[h!]
\centering
    \includegraphics[width=7cm]{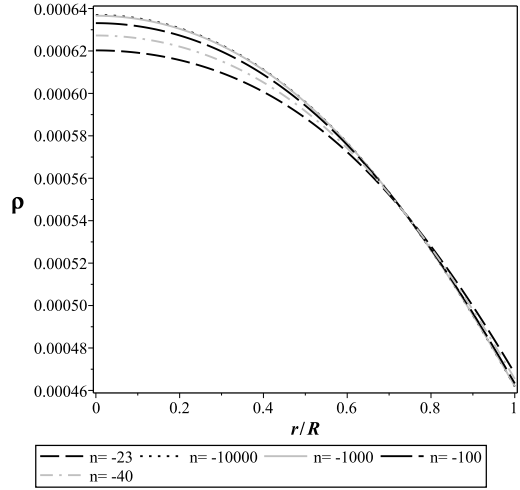}
\caption{Variation of density, $\rho$, with the fractional coordinate $r/R$ for $LMC\,X-4$. }
    \label{Fig3}
\end{figure}

From Eqs. (\ref{eq11}) and (\ref{eq12}), the anisotropy factor, $\Delta=p_t - p_r$, is given by

\begin{eqnarray}\label{eq14}
&\qquad\hspace{-2cm}\Delta= {\frac { \left[D\, \left( 1-A{r}^{2} \right) ^{n
}+ \left( n-2 \right)  \left( 1-A{r}^{2} \right)  \right]  \left[
  D\,\left( 1-A{r}^{2} \right) ^{n}+n \left( 1-A{r}^{2}
 \right)  \right] {A}^{2}{r}^{2}}{8\,\pi \, \left[ A\,D\,{r}
^{2} \left( 1-A{r}^{2} \right) ^{n}+ \left(1-A{r}^{2} \right) ^{2}
 \right] ^{2}}}.\nonumber \\
 \end{eqnarray}

\begin{figure}[h!]
\centering
    \includegraphics[width=7cm]{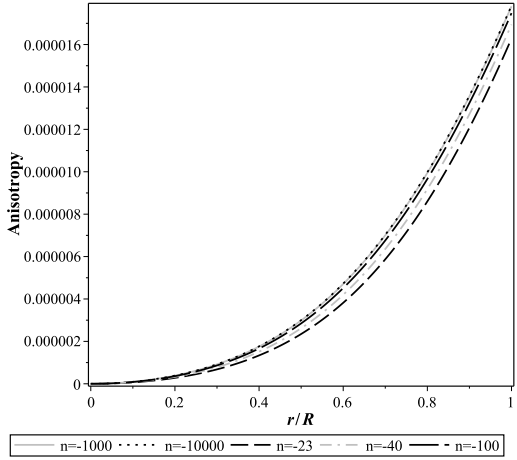}
\caption{Variation of the anisotropic factor ($\Delta$) with the fractional coordinate $r/R$ for $LMC\,X-4$. }\label{Fig4}
\end{figure}

One may find from Fig.~\ref{Fig4} that the anisotropy is zero at the center and a maximum at the surface, which is the same as the proposal of Deb et al.~\cite{Deb2016a}; for all anisotropic compact stars, the anisotropy should be maximal at the surface. The respective masses and radii of the strange star LMCX-4 are given in Table~1.  To plot the graphs, values of $A$, $B$, $K$, and $C$ were determined for different values of $n$.  These values are also listed in Table 2.

\section{Boundary conditions}\label{sec5}
The following boundary conditions must be satisfied for any physically acceptable anisotropic solution:

\subsection{(i)} The interior of metric (\ref{eq5}) for the anisotropic matter distribution should be matched at the boundary [$r=R$ of the star] with the exterior Schwarzschild metric, given by
\begin{eqnarray}\label{15}
&\qquad\qquad\hspace{-2.7cm}ds^{2} =\left(1-\frac{2M}{r} \right)\, dt^{2} -\left(1-\frac{2M}{r}
\right)^{-1} dr^{2}-r^{2} (d\theta ^{2}
+\sin ^{2} \theta \, d\phi ^{2} ),\nonumber \\
\end{eqnarray}
where $M$ is a constant representing the total mass of the compact star.

\subsection{(ii)} The radial pressure $p_{r}$ must be finite and positive
inside the stars, singularity free at the centre $r=0$, and it
should vanish at the boundary $r = R$ of the star~\cite{Misner1964}.

The radial pressure at the boundary $p_r(R) =0$ is
\begin{eqnarray}
&\qquad\hspace{1cm}D=\frac{2n(AR^2-1)}{(1-AR^2)^n}. \label{16}
\end{eqnarray}

The condition $e^{\nu(R)}=e^{-\lambda(R)}$ gives the value of the constant $B$ as
\begin{eqnarray}
&\qquad\hspace{-0.2cm}B=\frac{1}{(1-AR^2)^n\,[1+D\,AR^2\,(1-AR^2)^{n-2}]}.  \label{17}
\end{eqnarray}

The mass $M$ of the star is determined by using the condition $e^{-\lambda(R)}=1-\frac{2M}{R}$ and is given by
\begin{eqnarray}
&\qquad\hspace{1cm}\frac{M}{R}=\frac{D\,AR^2\,(1-AR^2)^{n-2}}{2\,[1+D\,AR^2\,(1-AR^2)^{n-2}]}.    \label{18}
\end{eqnarray}

Denoting the density of stars at the surface by ${{\rho}_{s}}$, the value of the constant $A$ is given by
\begin{eqnarray}\label{19}
&\qquad\hspace{-2cm} A={\frac {8\,\rho_{{s}}\pi \,{R}^{2}-3\,n+16\,\rho_{{s}}\pi \,{R}
^{2}n-\sqrt {-64\,\rho_{{s}}\pi \,{R}^{2}n+9\,{n}^{2}-32\,\rho_{{s}}
\pi \,{R}^{2}{n}^{2}}}{2\,{R}^{2} \left( -4\,{n}^{2}+n+16\,\rho_{{s}}\pi
\,{R}^{2}{n}^{2}+4\,\rho_{{s}}\pi \,{R}^{2}+16\,\rho_{{s}}\pi \,{R}^{2
}n \right) }}.\nonumber \\
\end{eqnarray}

\section{physical properties of the anisotropic solution}\label{sec6}

\subsection{Regularity at the center}\label{subsec1}
(i) The metric potentials at the center ($r=0$) are: $e^{\nu(0)}=B$ and $e^{-\lambda(0)}=1$; this shows that metric potentials are finite and singularity free at the center ($r=0$). The behavior can be seen inside the star in Fig.~\ref{Fig1}.\\

(ii) Radial pressure at the center: $p_r(0)=-\frac{A\,(2n+D)}{8\pi}$, tangential pressure at the center: $p_t(0)=-\frac{A\,(2n+D)}{8\pi}$. Since $p_r$ and $p_t$ are positive and finite at the center (Fig.~\ref{Fig2}), this implies that
\begin{equation}
D < -2\,n.  \label{eq20}
\end{equation}

(iii) Matter density at the center: $\rho(0)=\frac{3\,A\,D}{8\,\pi}$. Since the matter density is positive at the center (Fig.~\ref{Fig3}) and $A$ is also positive, we have
\begin{equation}
  D > 0. \label{eq21}
\end{equation}

(iv) For physically acceptable models, the pressure should be dominated by the matter density throughout the interior of the star, i.e., 
$\omega_r=p_r/\rho <1$ and $\omega_t=p_t/\rho < 1$. This gives
\begin{equation}
  -\frac{n}{2} < D.  \label{eq22}
\end{equation}

Hence the conditions (ii), (iii) and (iv) will be valid simultaneously only with the choice of negative values of $n$, and this will yield a positive value for $D$. Using Eqs.~(\ref{eq20})-(\ref{eq22}), we get the inequality
\begin{equation}
  -\frac{n}{2} < D < -2\,n  \label{eq23},
\end{equation}
where $n<0$. Tables 1 and~2 show that $D$ maintains this inequality.

From our model we find the radial ($\omega_r$) and tangential ($\omega_t$) EOS parameters as
\begin{eqnarray}\label{eq24}
&\qquad\hspace{-2cm}\omega_r={\frac { \left[ 2\,n \left( A{r}^{2}-1 \right) -  D
\left( 1-A{r}^{2} \right) ^{n} \right]  \left[  \left( 1-A{r}^{2}
 \right) ^{2}+  A\,D\,{r}^{2} \left( 1-A{r}^{2} \right) ^{n
} \right] }{  D \left( 1-A{r}^{2} \right) ^{n} \left[ 3
-2\, \left( n+1 \right) A{r}^{2}+ A\,D\,{r}^{2} \left( 1-A
{r}^{2} \right) ^{n}+ \left( 2\,n-1 \right) {A}^{2}{r}^{4} \right] }}, \nonumber \\  \\
\label{eq25}
&\qquad\hspace{-2cm}\omega_t= {\frac { \left( A{r}^{2}-1 \right)  \left[ 2\,n \left( 1-A{r}^{2}
 \right) +{n}^{2}A{r}^{2} \left( A{r}^{2}-1 \right) + D\,
 \left( 1+A{r}^{2} \right)  \left( 1-A{r}^{2} \right) ^{n}
 \right] }{  D \left( 1-A{r}^{2} \right) ^{n} \left[ 3
-2\, \left( n+1 \right) A{r}^{2}+ A\,D\,{r}^{2} \left( 1-A
{r}^{2} \right) ^{n}+ \left( 2\,n-1 \right) {A}^{2}{r}^{4} \right] }}.\nonumber \\
\end{eqnarray}

\begin{figure} [h!]
\centering
    \includegraphics[width=6.7cm]{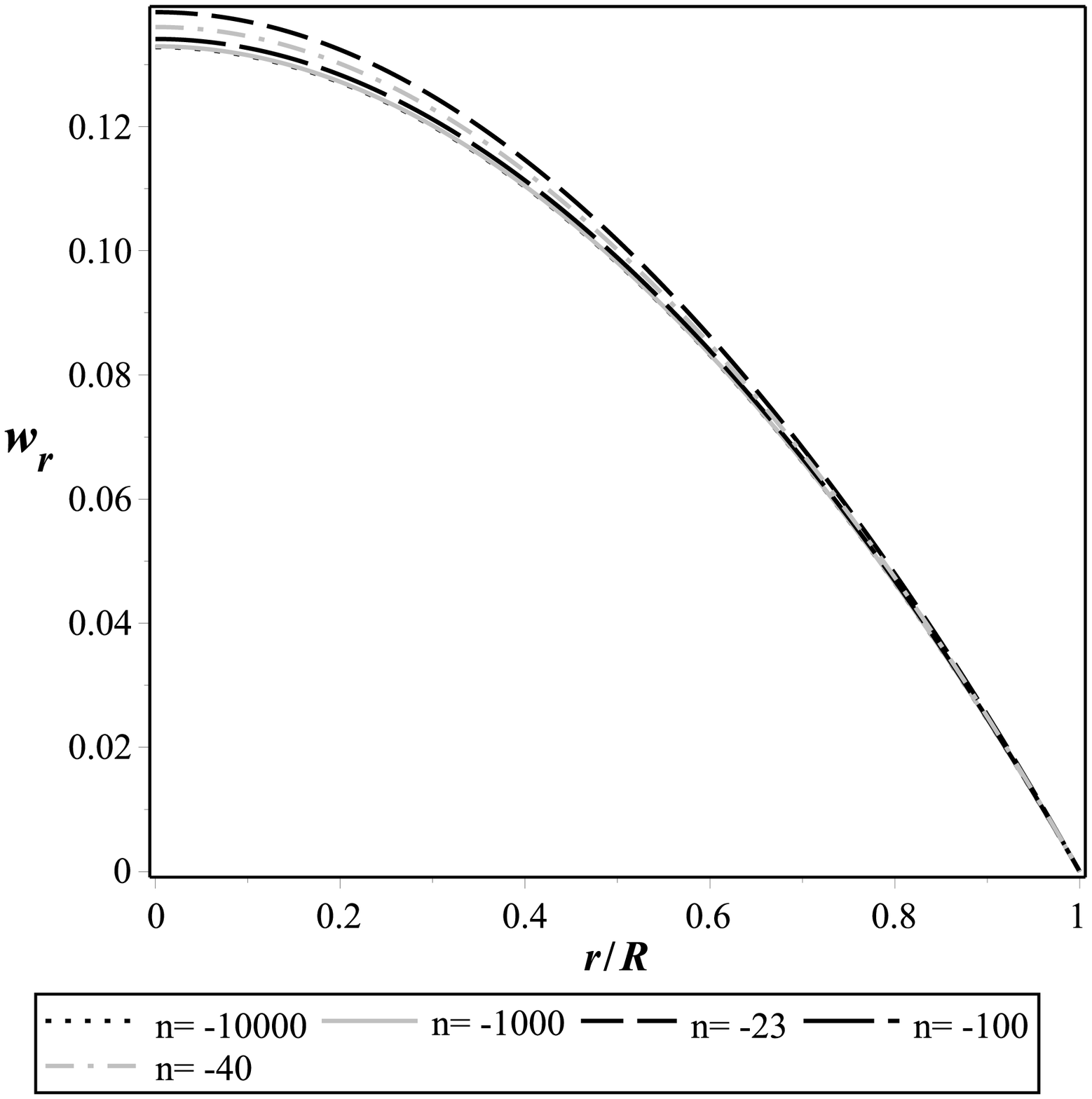}
    \includegraphics[width=6.7cm]{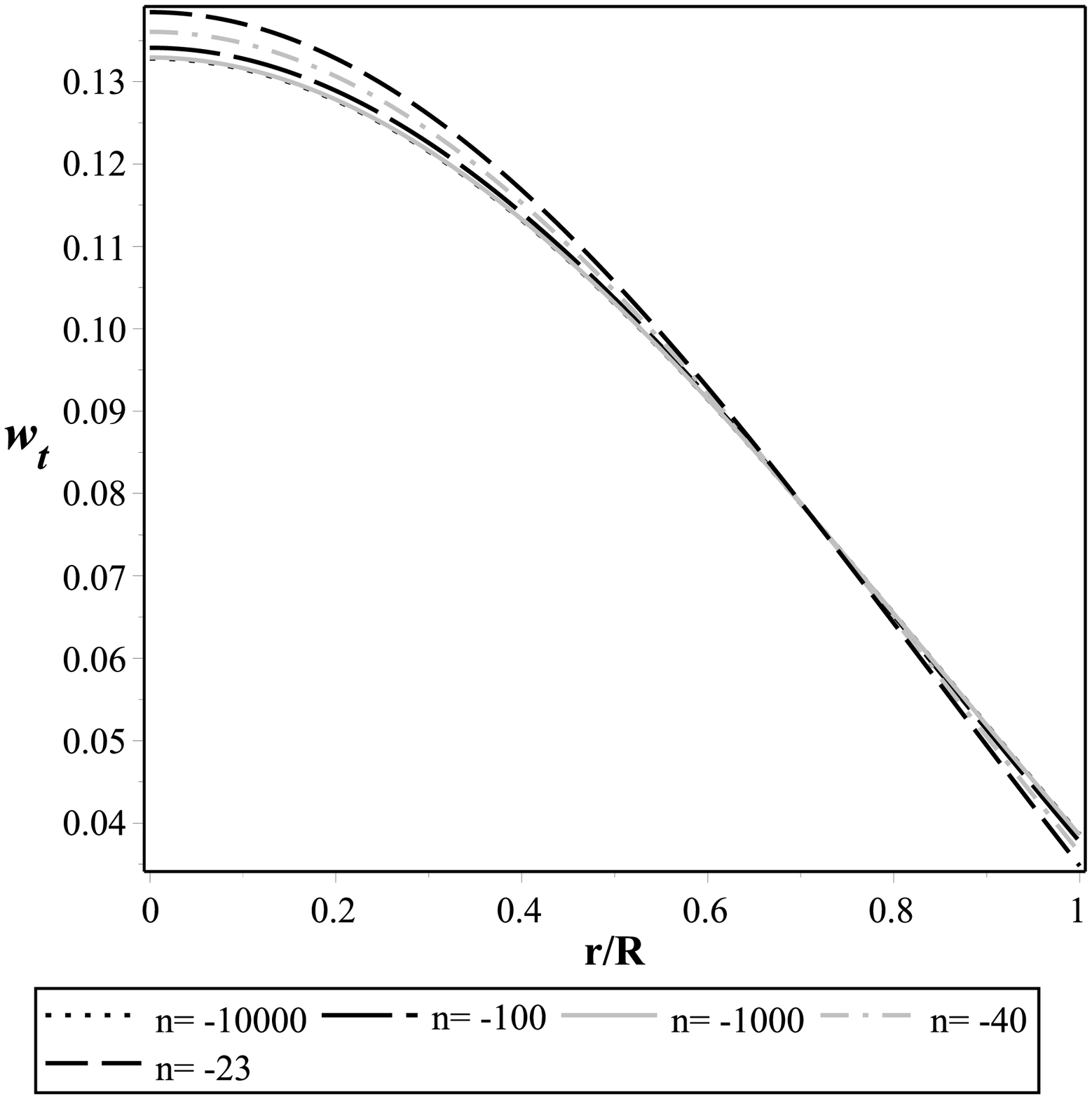}
\caption{Variation of the parameters ${\omega}_r$ (left panel) and ${\omega}_t$ (right panel) with respect to the fractional coordinate $\frac{r}{R}$ for $LMC\,X-4$. } \label{Fig5}
\end{figure}

The behavior of $\omega_r$ and $\omega_t$ can be seen from Fig.~\ref{Fig5}. From Fig.~\ref{Fig5}, it can be observed that the pressures $p_r$ and $p_t$ are dominated by the density throughout the star since the functions $\omega_{r}$ and $\omega_{t}$ lie between 0 and 1. Also, we can see from this figure that both functions $\omega_{r}$ and $\omega_{t}$ are monotonically decreasing with increasing $r$. This implies that the temperature of the anisotropic compact star models decreases outward from the center.

\subsection{Energy condition}\label{subsec3}
The anisotropic stellar configuration will satisfy the usual energy conditions; the null energy condition (NEC), the weak energy
condition (WEC) and the strong energy condition (SEC) if the following inequalities are met for all the stellar models:
\begin{eqnarray}\label{eq27}
&\qquad\hspace{-1cm} NEC: \rho\geq 0,\\ \label{eq28}
&\qquad\qquad\hspace{-2.5cm} WEC: \rho+ {{p}_{r}} \geq  0 \hspace{0.1cm}({WEC}_{r}) \hspace{0.2cm} and
     \hspace{0.2cm} \rho + p_t \geq  0 \hspace{0.1cm}({WEC}_{t}),\nonumber \\ \\ \label{eq29}
&\qquad\hspace{-1cm} SEC: \rho+p_r+2{{p}_{t}} \geq  0.
\end{eqnarray}

\begin{figure}[h!]
\centering
	\includegraphics[width=6.7cm]{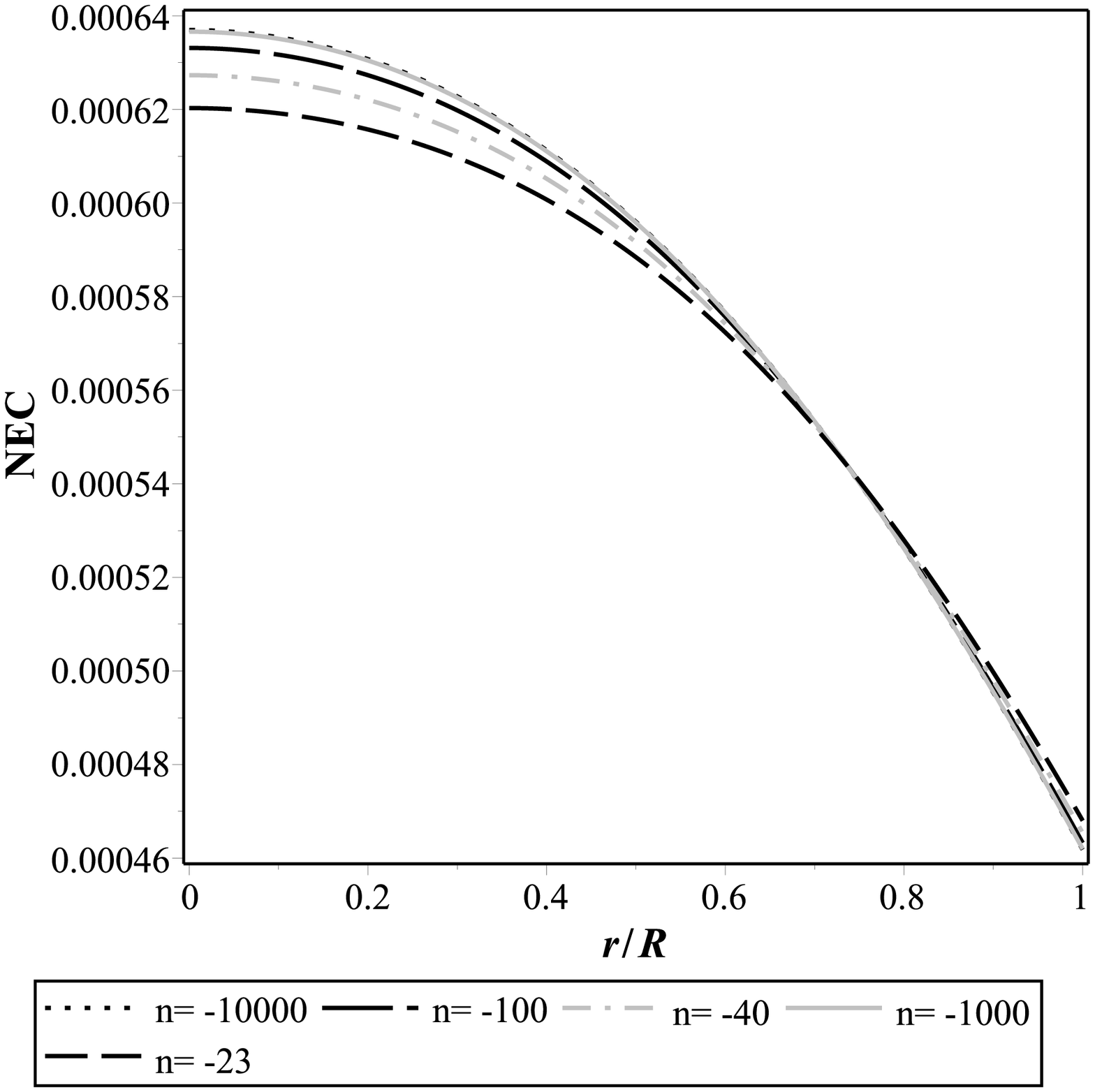}
    \includegraphics[width=6.7cm]{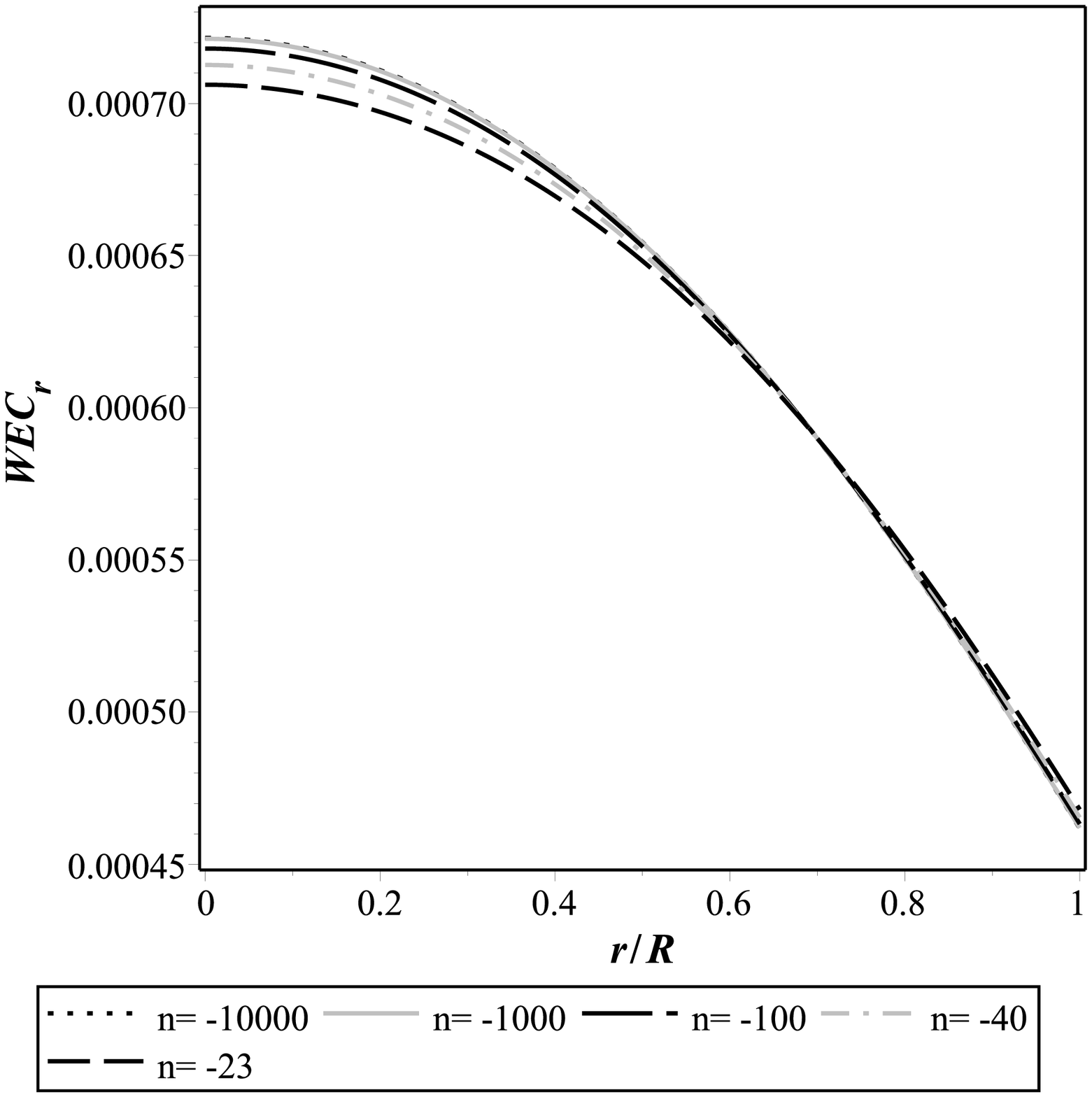}
     \includegraphics[width=6.7cm]{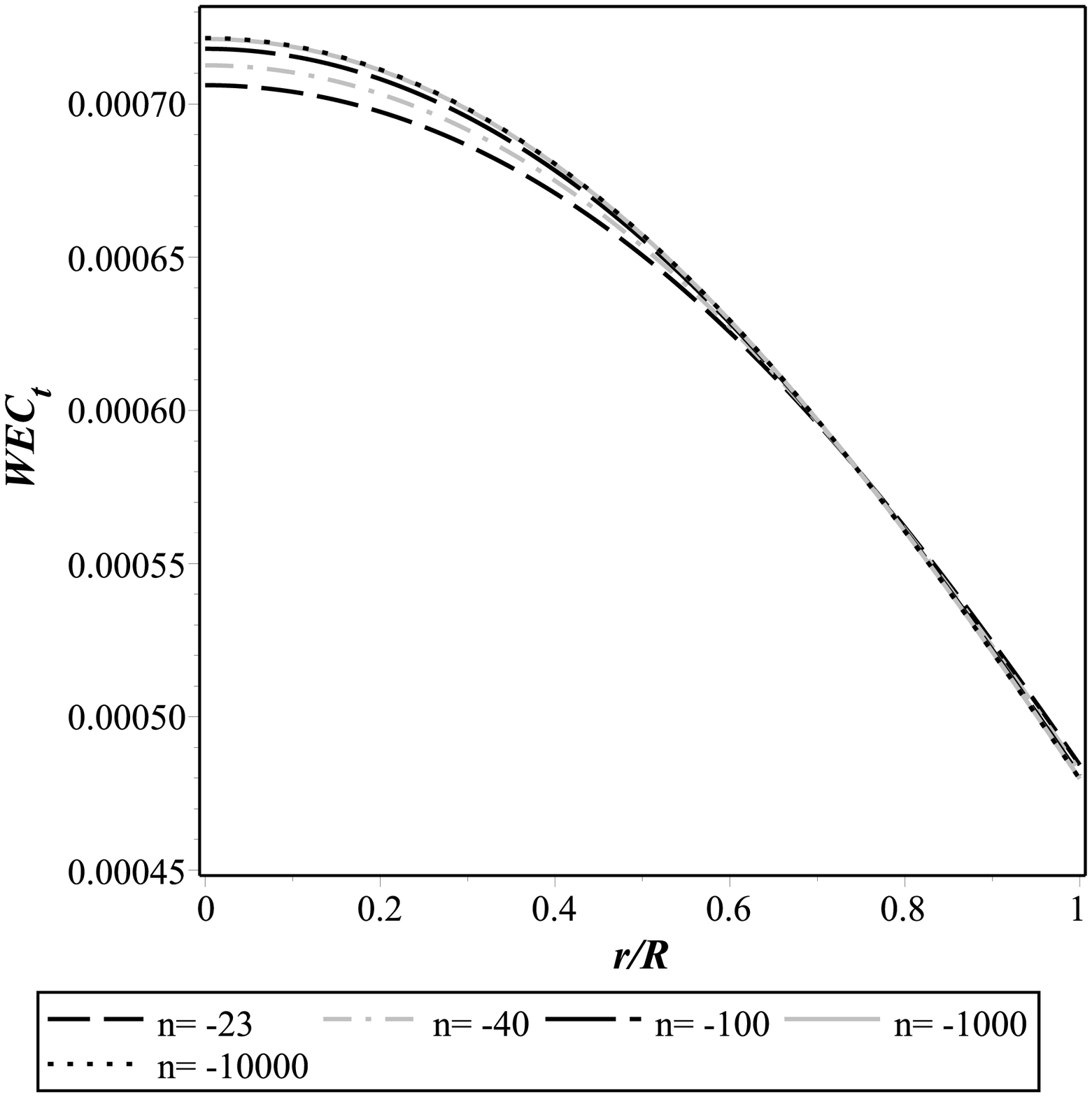}
     \includegraphics[width=6.7cm]{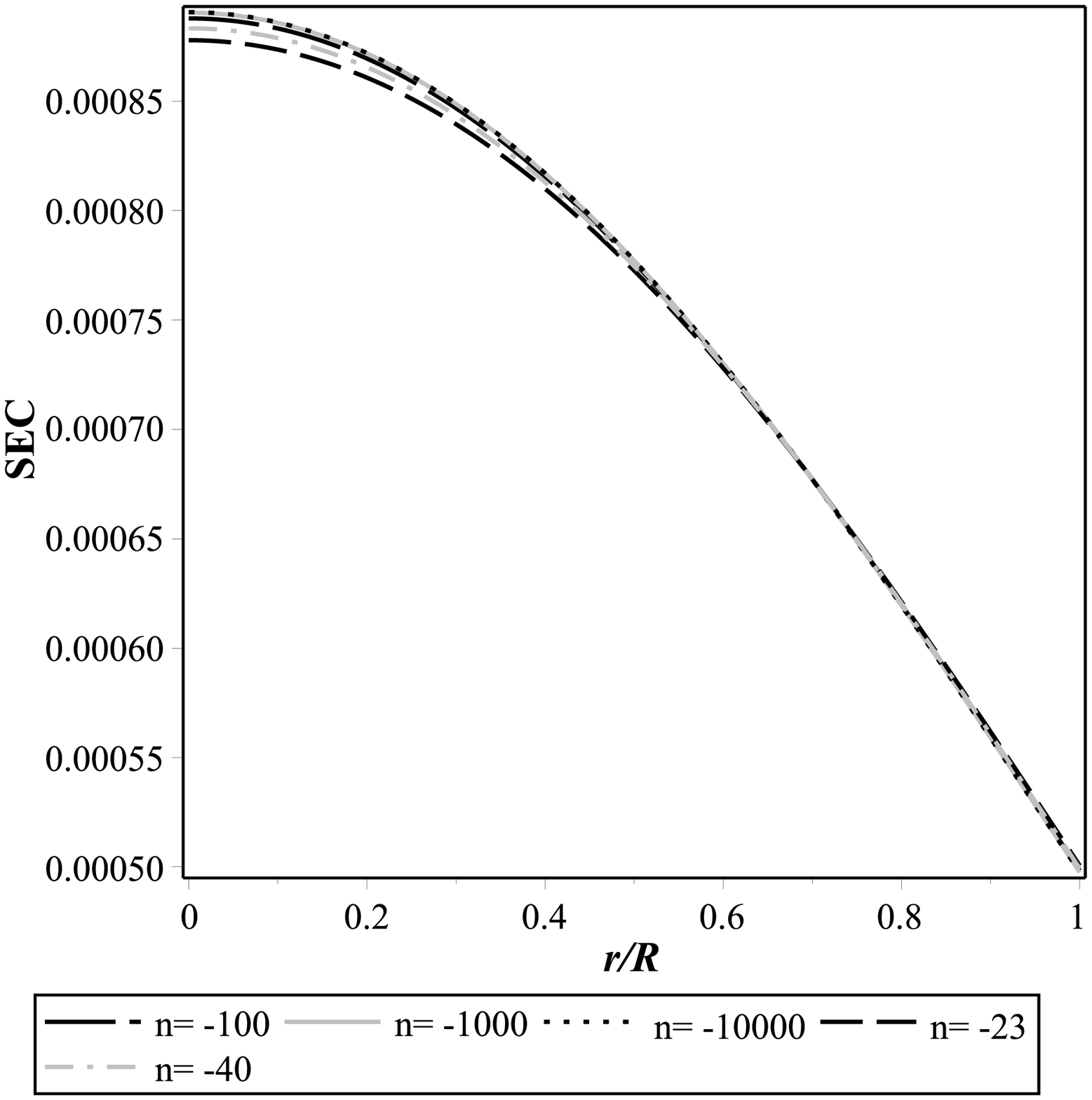}
	\caption{Variation of the energy conditions with respect to the fractional radius ($r/R$) for $LMC\,X-4$: (i) NEC (top left), (ii) $WEC_r$ for (top right), (iii) $WEC_t$ (bottom left), (iv) SEC (bottom right). }
\label{Fig7}
\end{figure}

From Fig.~\ref{Fig7} one may observe that all the energy conditions are met for the proposed anisotropic stellar model.

\subsection{Stability of the models}\label{subsec4}

\subsubsection{Stability under three different forces}\label{subsubsec1}
Tolman~\cite{Tolman1939} and Oppenheimer and Volkoff~\cite{Oppenheimer1939} predicted that for a stable stellar fluid sphere the sum of the forces, viz. gravitational force ($F_g$), hydrostatics force ($F_h$) and anisotropic force ($F_a$) must be zero for the system to be in equilibrium, i.e.,
\begin{equation}
F_g+F_h+F_a=0. \label{eq36}
\end{equation}

Following the works of Varela et al.~\cite{Varela2010} and Ponce de Le{\'o}n~\cite{Leon1993}, we use a generalized form of the Tolman-Oppenheimer-Volkoff (TOV) equation for our system, given by
\begin{eqnarray}
& \qquad\hspace{-1cm}-\frac{M_G(r)(\rho+p_r)}{r^2}e^{\frac{\lambda-\nu}{2}}-\frac{dp_r}{dr}+\frac{2}{r}(p_t-p_r)=0, \label{eq30}
\end{eqnarray}
where $M_G(r)$ is the gravitational mass of the stellar configuration. Now we have from the Tolman-Whittaker formula~\cite{Devitt1989} and the Einstein field equations $M_G(r)$ the form
\begin{eqnarray}
&\qquad\hspace{1cm} M_G(r)=\frac{1}{2}{{r}^{2}}e^{\frac{\nu-\lambda}{2}}\nu'. \label{eq31}
\end{eqnarray}

Now from Eqs.~(\ref{eq30}) and (\ref{eq31}) we have
\begin{eqnarray}
&\qquad\hspace{-1cm}-\frac{\nu'}{2}(\rho+p_r)-\frac{dp_r}{dr}+\frac{2}{r}(p_t-p_r)=0, \label{eq32}
\end{eqnarray}
where the three terms in Eq.~(\ref{eq32}) represents three forces $F_g$, $F_h$ and $F_a$, respectively.
For our system the explicit forms of the forces are
\begin{eqnarray}\label{eq37}
&\qquad\hspace{-7cm} F_g=-\frac{\nu'}{2}(\rho+p_r)\nonumber \\
&\qquad\hspace{-2cm}=-\frac{n\,A^2r}{4\pi}\,\left[\frac{-D(1-Ar^2)^n\,(1+Ar^2)+n\,(1-Ar^2)^2+F_{{{\it a1}}}}{[(1-Ar^2)^2+D\,Ar^2\,(1-Ar^2)^n]^2}\right],\nonumber \\ \\ \label{eq38}
&\qquad\hspace{-8cm} F_h=-\frac{dp_r}{dr}\nonumber \\
&\qquad\hspace{-1cm}=-\frac{A^2r}{4\pi}\,\left[\frac{-n[D(1-Ar^2)^n\,(Ar^2-3)+2\,(1-Ar^2)^2]-n\,F_{{{\it a1}}}+F_{h1}}{[(1-Ar^2)^2+D\,Ar^2\,(1-Ar^2)^n]^2}\right],\nonumber \\ \\ \label{eq39}
&\qquad\hspace{-7cm} F_a=\frac{2}{r}(p_t-p_r)\nonumber \\
&\qquad\hspace{-1cm}=\frac{A^2r}{4\pi}\,\left[\frac{n^2\,(1-Ar^2)^2-2n(1-Ar^2)^2+2{D}n(1-Ar^2)^{n+1}+F_{h1}}{[(1-Ar^2)^2+D\,Ar^2\,(1-Ar^2)^n]^2}\right],\nonumber \\ 
\end{eqnarray}
 where $F_{{{\it a1}}}=2\,n  D \, A{r}^{2} \left(1 -A{r}^{2} \right) ^{n}$, $F_{h1}=D(1-Ar^2)^n[-2(1-Ar^2)+D(1-Ar^2)^n]$.

\begin{figure}[h!]
\centering
\includegraphics[width=5.5cm]{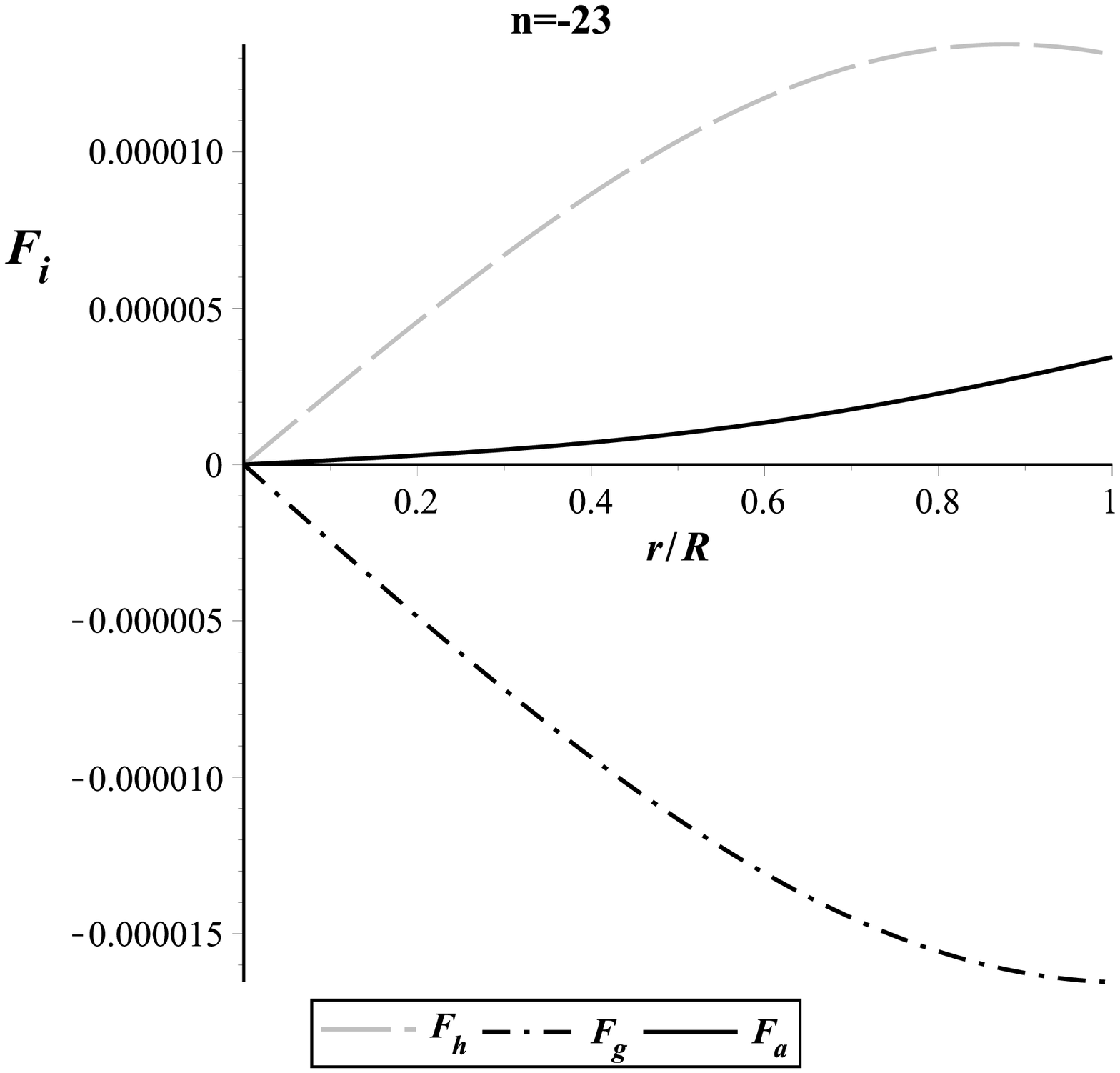}
\includegraphics[width=5.5cm]{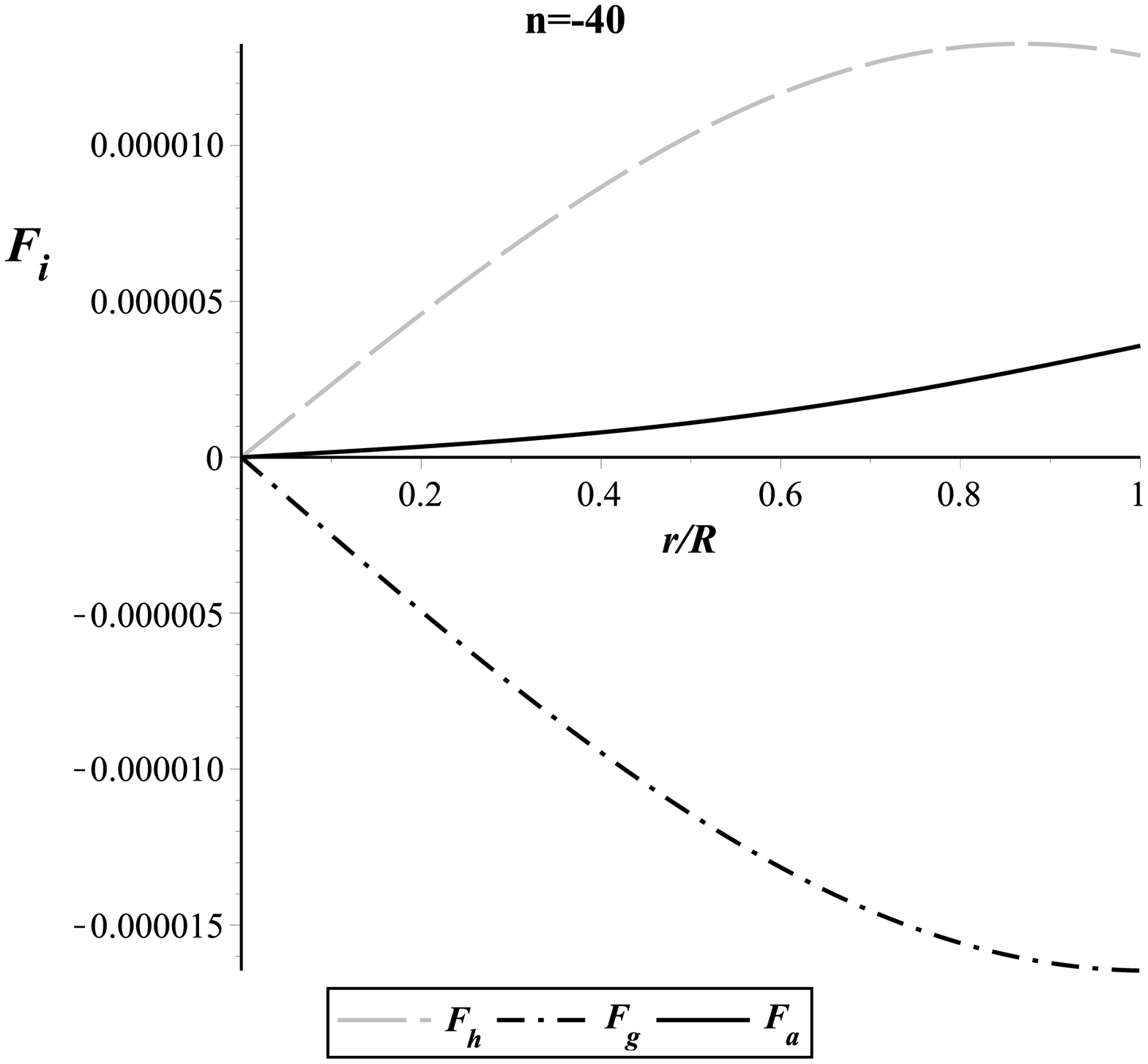}
\includegraphics[width=5.5cm]{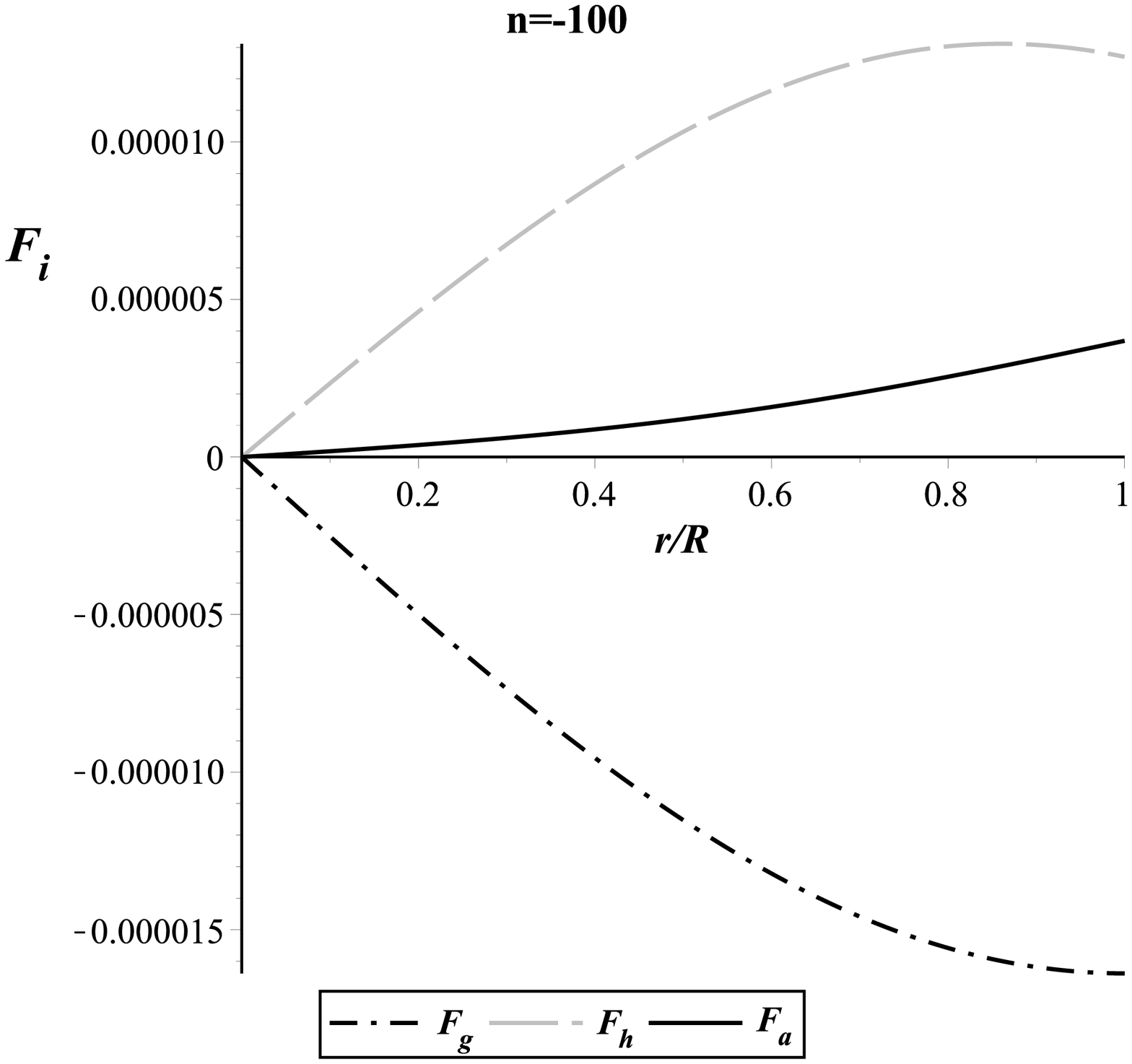}
\includegraphics[width=5.5cm]{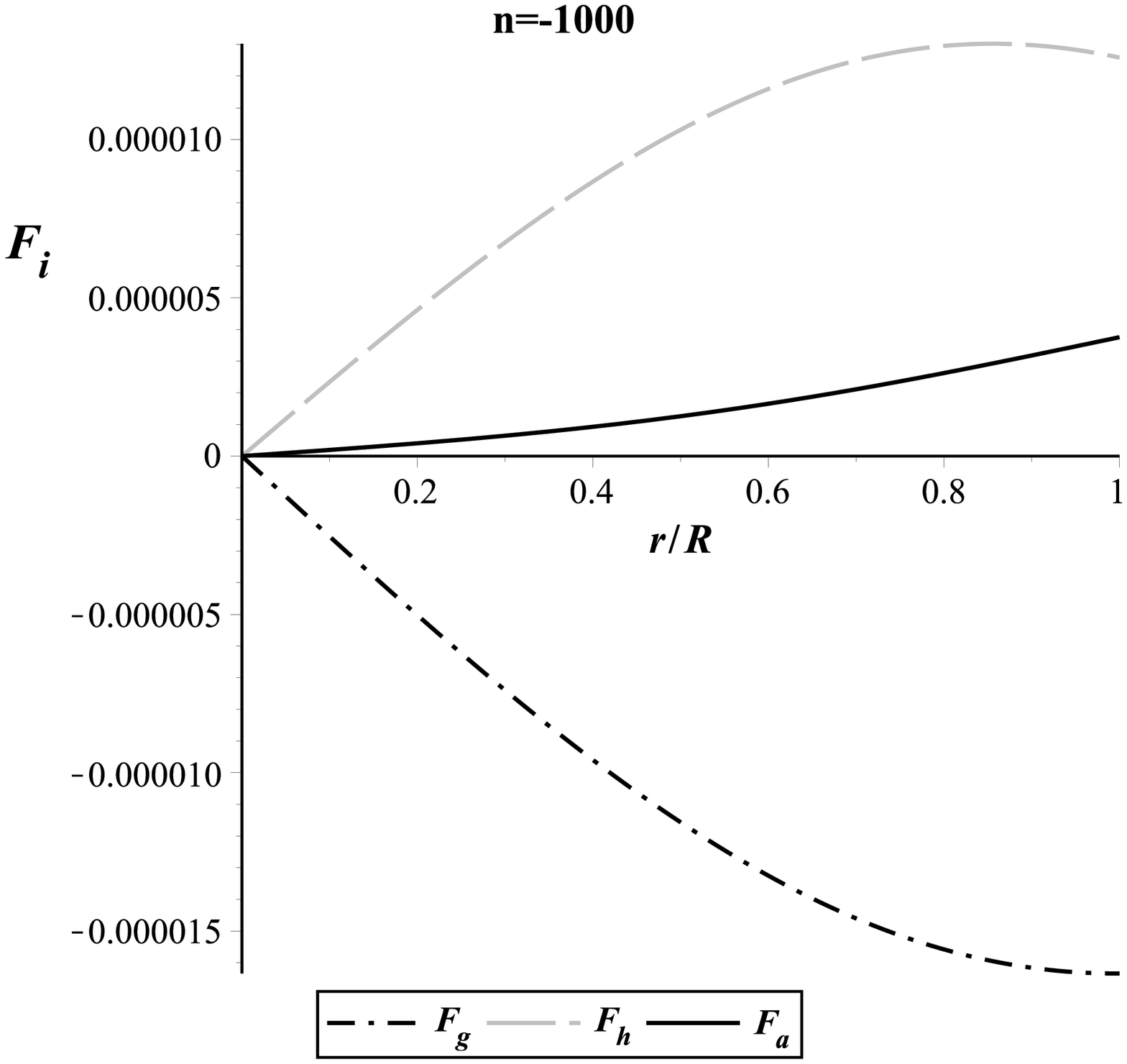}
\includegraphics[width=6cm]{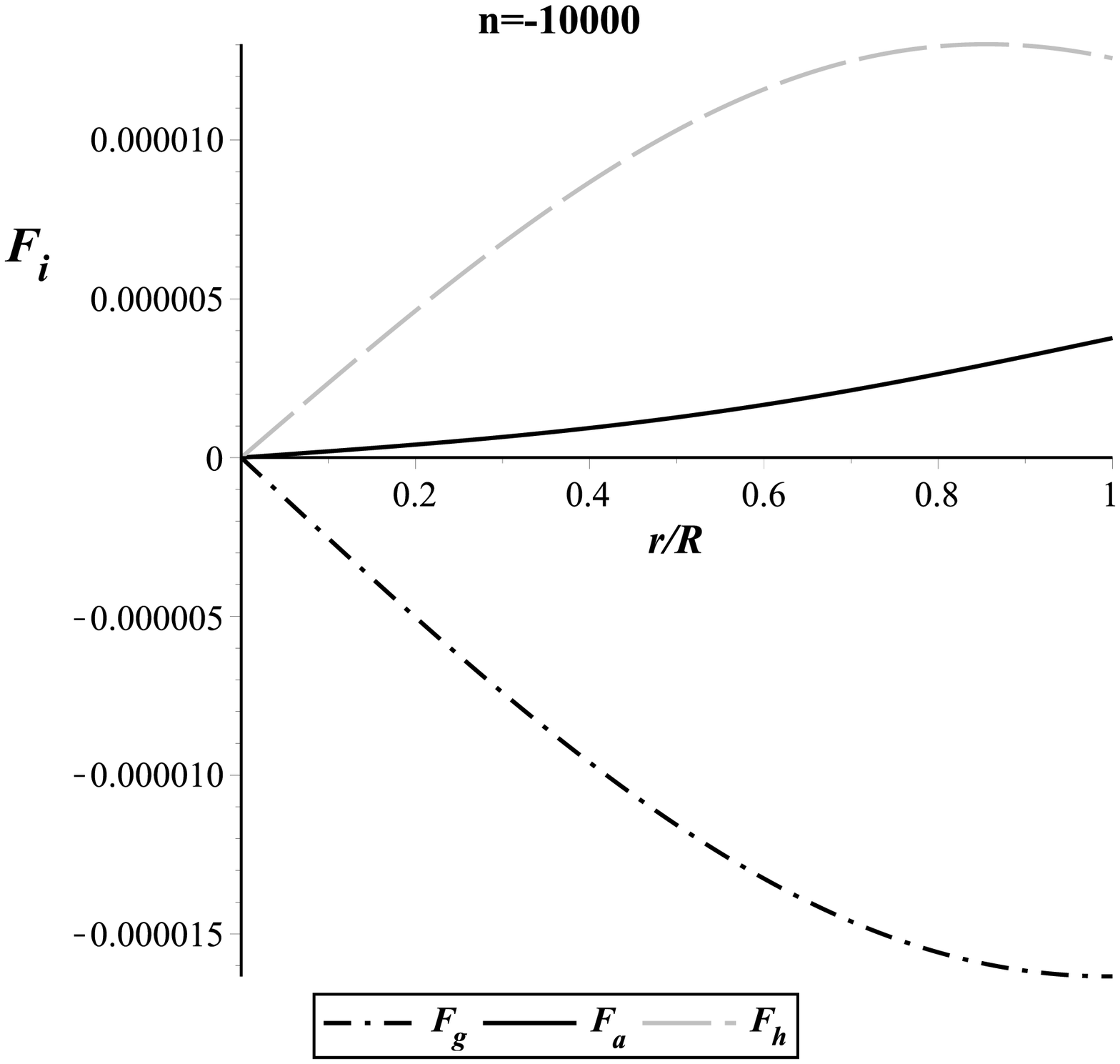}
	\caption{Variation of the different forces with respect to the fractional radius $r/R$ for $LMC\,X-4$. }
	\label{Fig8}
\end{figure}

 Fig.~\ref{Fig8} features the behaviour of the TOV equation for the star $LMC\,X-4$ for different values of $n$. It is also clear from Fig.~\ref{Fig8} that our system is in equilibrium.

\subsubsection{Herrera cracking concept}\label{subsubsec2}
To test the stability of the system, we also study the `cracking concept' of Herrera's~\cite{Herrera1992}, which predicts that for the stability of the stellar configuration, the following conditions need to be valid for the system:\\

i) Condition of causality: inside the stellar configuration the square of the radial ($v^2_r$) and tangential ($v^2_t$) must lie within the limit $0$ to $1$, so that the velocity of sound could not exceed the velocity of light, thereby maintaining the physical validity of the system.

ii) No cracking: Herrera~\cite{Herrera1992} and Abreu et al.~\cite{Abreu2007} predicted that for a stable region the square of the radial sound speed should be greater than the square of the tangential sound speed throughout the region and must maintain the inequality        
$|v^2_t- v^2_r| \leq 1$.

For our system, the square of the sound velocities take the following forms:
\begin{eqnarray}\label{33}
&\qquad\hspace{-7cm} v^2_r=\frac{dp_r}{d\rho}\nonumber \\ 
&\qquad\hspace{-1cm}=\frac{{[(1-Ar^2)^2+D\,Ar^2\,(1-Ar^2)^n]}\left[-n\,D(1-Ar^2)^n\,(Ar^2-3)+p_{r1}\right]}{-D\,(1-Ar^2)^n\left(\rho_1+\rho_2+\rho_3\right)}, \\ 
\label{34}
&\qquad\hspace{-3cm} v^2_t=\frac{dp_t}{d\rho}=\frac{\left(p_{t1}+p_{t2}+p_{t3}\right)}{-\,D\,(1-Ar^2)^n\left(\rho_1+\rho_2+\rho_3\right)},
\end{eqnarray}
where\\
\indent $p_{r1}=\big[D(1-Ar^2)^{n+1}\big\lbrace-2+D(1-Ar^2)^{n-1}\big\rbrace-2\,n(1-Ar^2)^2\big\lbrace1+\,D\,n\,Ar^2\,(1-Ar^2)^{n-2}\big\rbrace\big]$, 

 $p_{t1}=\big[2\,D\,n^3\,A^2r^4\,(1-Ar^2)^{n+1}+2D(1-Ar^2)^n \left\lbrace-2+D(1-Ar^2)^n+3Ar^2-A^3r^6\right\rbrace\big]$,

$p_{t2}=\big[n^2\,(1-Ar^2)\big\lbrace(1-Ar^2)-D\,Ar^2(1-Ar^2)^n\,(5+Ar^2)-A^2r^4+A^3r^6\big\rbrace\big]$,

$p_{t3}=\big[-n\big\lbrace4(1-Ar^2)^3+D^2\,(1-Ar^2)^{2n}\,Ar^2\,(1+Ar^2)-D\,(1-Ar^2)^{n+1}\,(5-A^2r^4)\big\rbrace\big]$,

$\rho_1=\big[D^2\,(1-AR^2)^{2n}\,Ar^2-2(1-Ar^2)^2\,(5+Ar^2)-2\,n^2\,Ar^2\,(1-Ar^2)^2\big]$,

$\rho_2=\big[2\,n^2\,D\,A^2r^4\,(1-Ar^2)^n+D\,(1-Ar^2)^{n}\,(5-4\,Ar^2+3\,A^2r^4)\big]$,

$\rho_3=\big[n\,(1+Ar^2)\,\big\lbrace5-10\,Ar^2-3D\,Ar^2\,(1-Ar^2)^n+5\,A^2r^4\big\rbrace\big]$.

\begin{figure}[h!]
\centering
    \includegraphics[width=6.7cm]{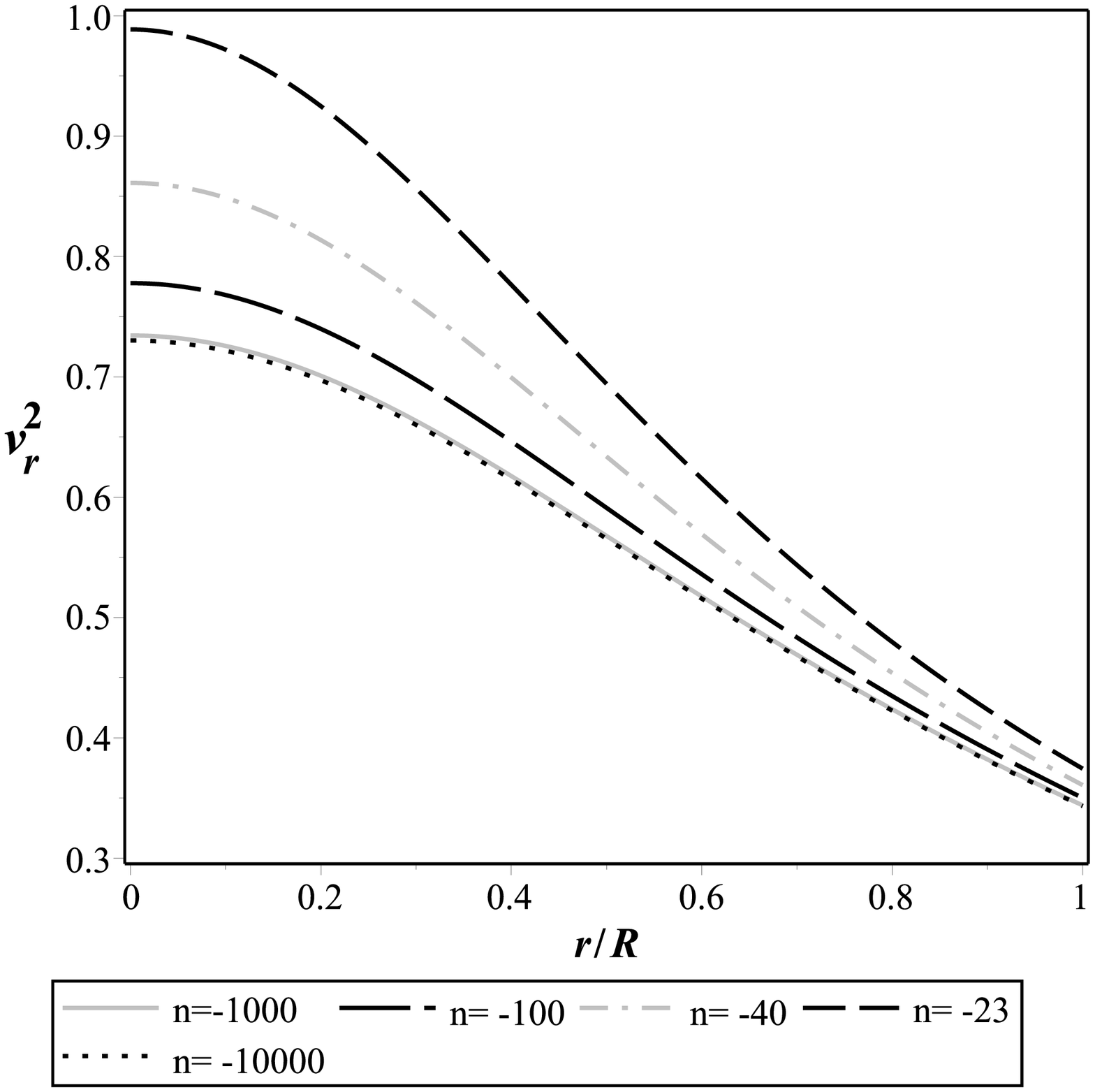}
    \includegraphics[width=6.7cm]{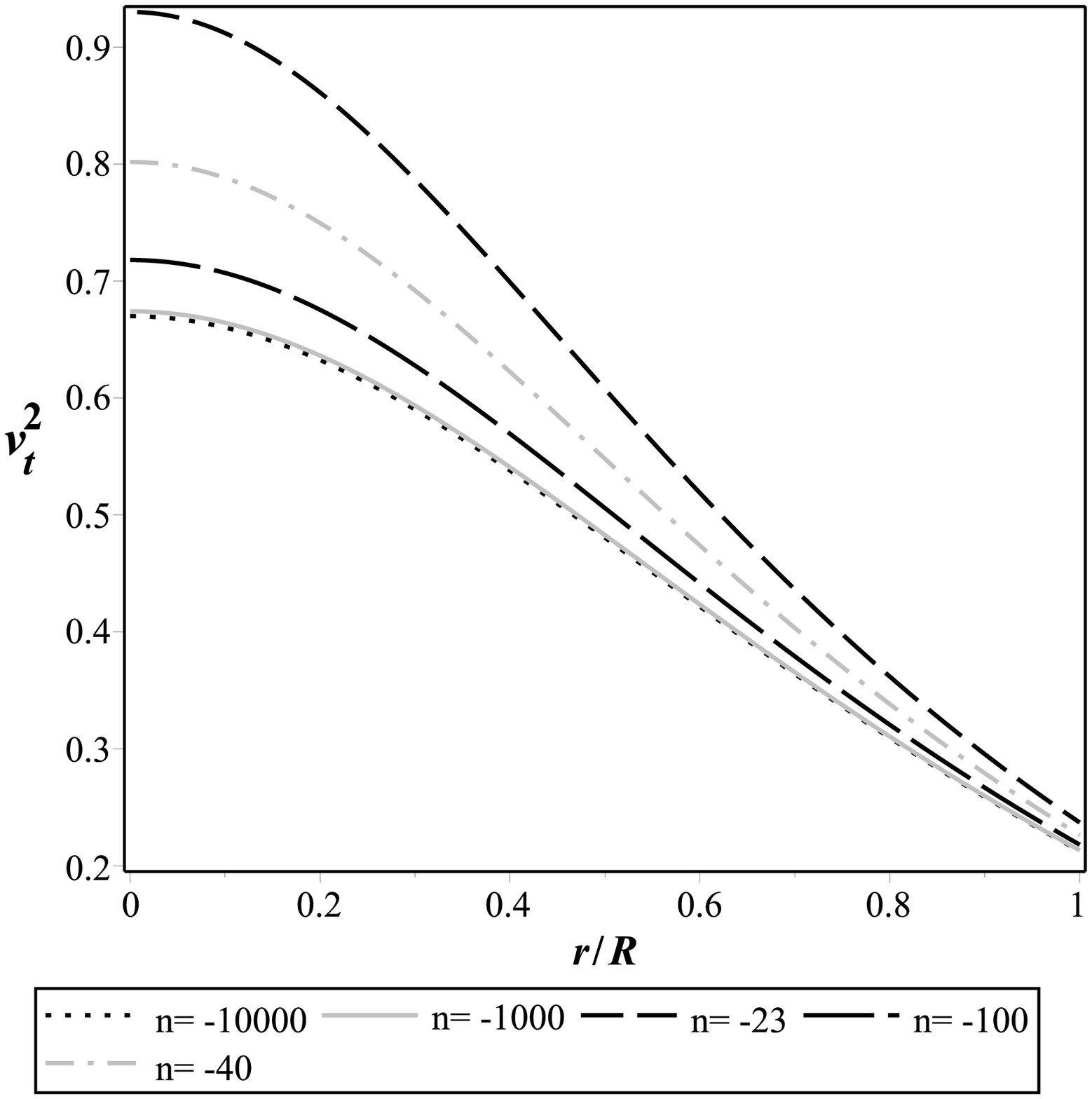}
\caption{Variation of the radial velocity $v^2_r$ (left panel) and the transverse velocity $v^2_t$  (right panel) with respect to the fractional coordinate $r/R$ for $LMC\,X-4$. }
    \label{Fig6}
\end{figure}

\begin{figure}[h!]
\centering
    \includegraphics[width=7cm]{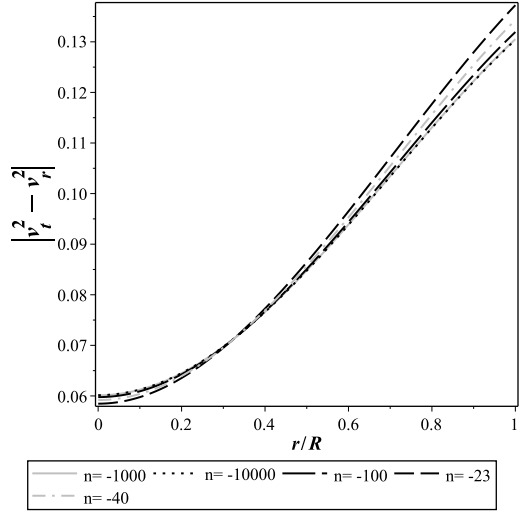}
\caption{Variation of the velocity difference $|v^2_t-v^2_r|$ with the radial coordinate $r/R$ for $LMC\,X-4$. }
    \label{Fig9}
\end{figure}

From Figs.~\ref{Fig6} and~\ref{Fig9} it is clear that both of the conditions, (i) the condition of causality and (ii) the no cracking condition, are satisfied within the anisotropic spherically symmetric stellar model.

\subsubsection{Adiabatic index}\label{subsubsec3}
Heintzmann and Hillebrandt~\cite{Heintzmann1975} predicted that for a stable isotropic stellar configuration, it is required that the adiabatic index, $\gamma$, should be greater than $4/3$ inside the stellar system. For our system the radial adiabatic index $(\Gamma_r)$ and the tangential adiabatic index $(\Gamma_t)$ are, repectively,
\begin{eqnarray}\label{eq40}
&\qquad\hspace{0cm}\Gamma_r=\frac{p_r+\rho}{p_r}\,\frac{dp_r}{d\rho}=\frac{p_r+\rho}{p_r}\,v^2_r, \\ \label{eq41}
&\qquad\hspace{0cm}\Gamma_t=\frac{p_t+\rho}{p_t}\,\frac{dp_t}{d\rho}=\frac{p_t+\rho}{p_t}\,v^2_t. 
\end{eqnarray}
\begin{figure}[h!]
\centering
    \includegraphics[width=6.7cm]{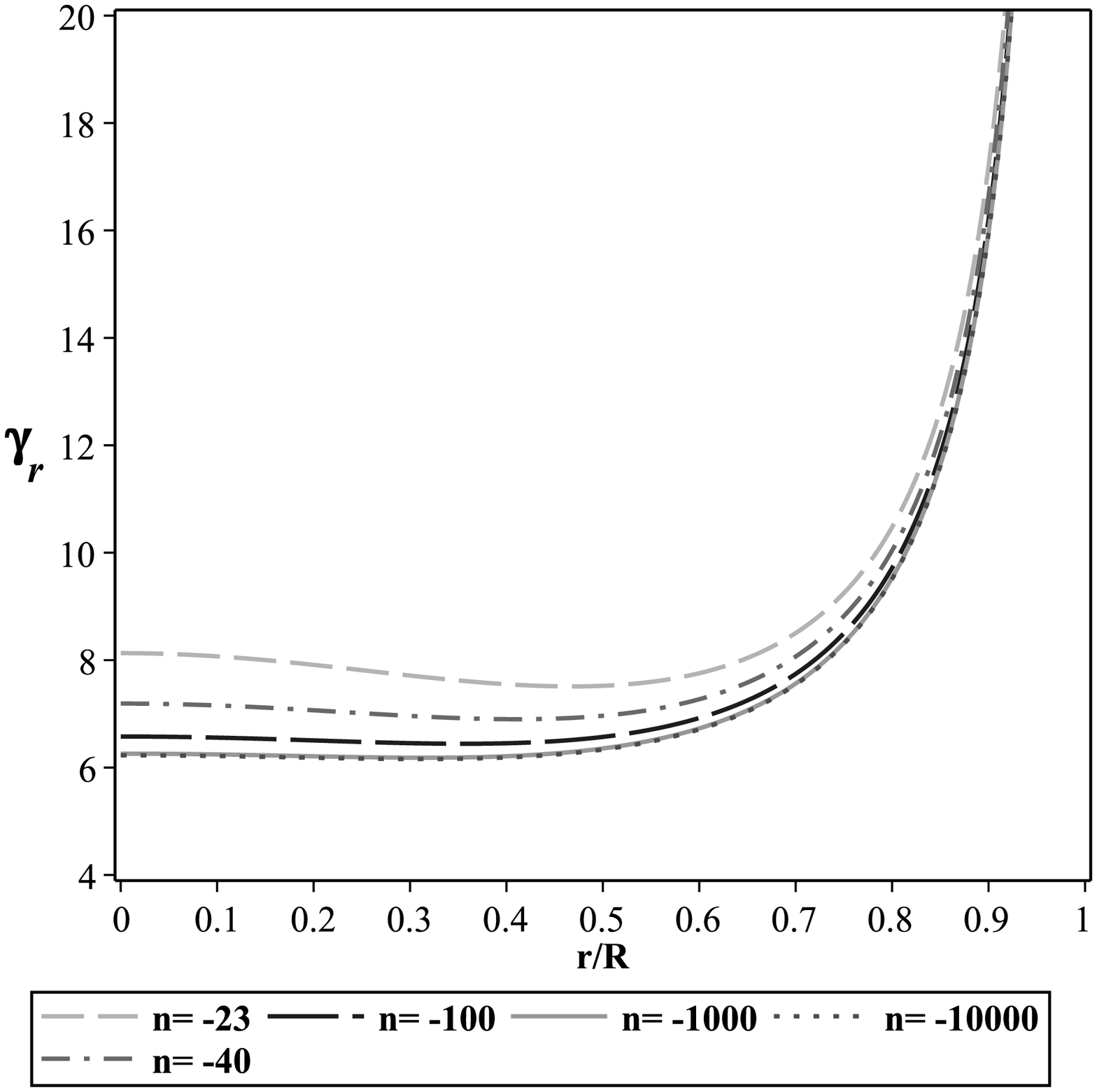}
    \includegraphics[width=6.7cm]{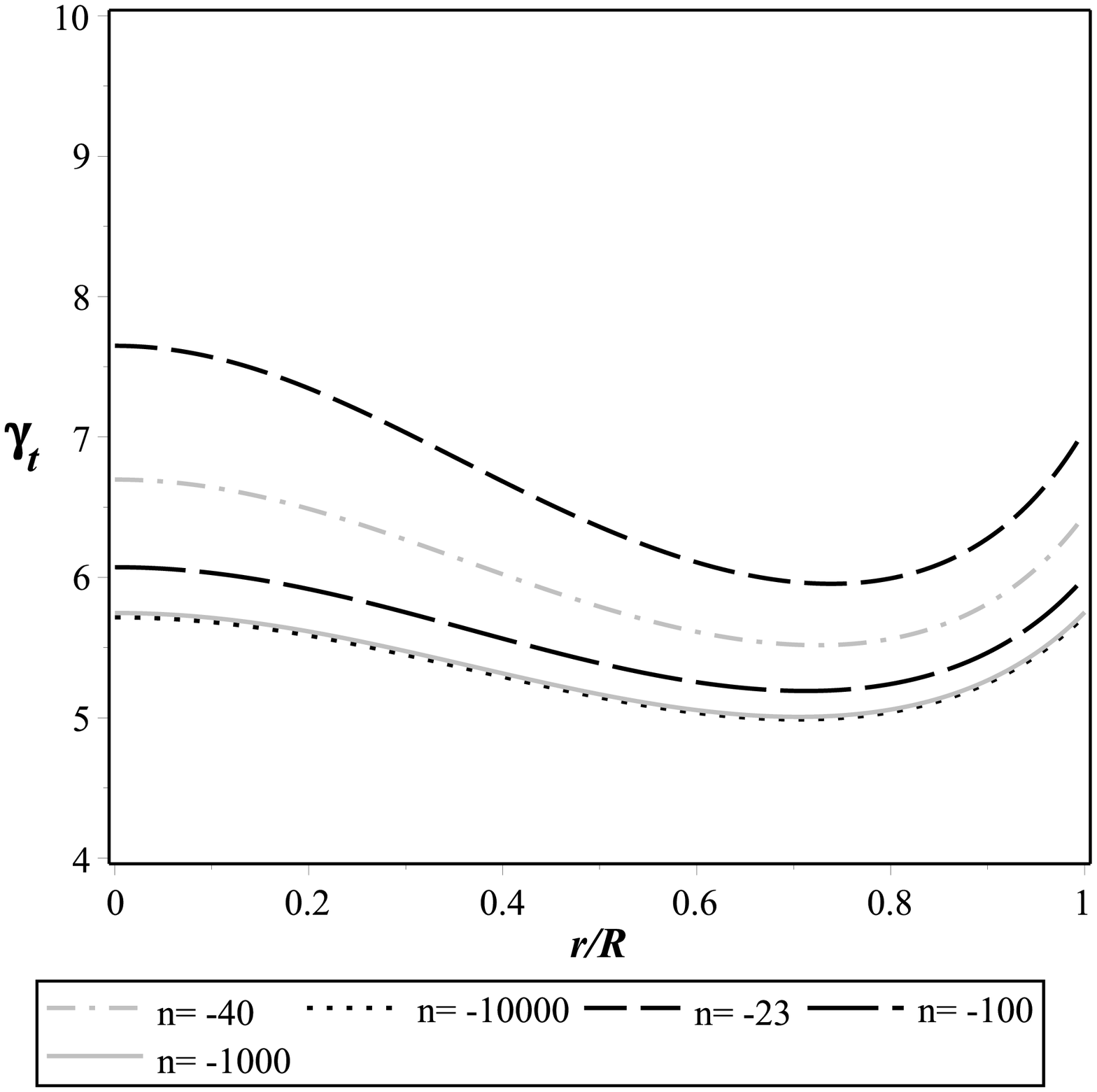}
\caption{Variation of the adiabatic index $\gamma_r$ (left panel) and $\gamma_t$ (right panel) with the radial coordinate ($r/R$) for $LMC\,X-4$. }
    \label{Fig10}
\end{figure}

From Fig.~\ref{Fig10} it is clear that the stellar system is stable under the condition that $\Gamma_r$ and $\Gamma_t$ are  greater than $4/3$ for the different values of $n$.

\subsection{Effective mass and compactness parameters}\label{subsec5}
Buchdahl~\cite{Buchdahl1959} proposed that for a spherically symmetric static stellar configuration the ratio of the maximum allowed mass to the radius is 4/9, usually stated as $2M/R \leq 8/9$, which was later proposed by Mak and Harko~\cite{Mak2003} in the more generalized form.

For our model the effective gravitational mass takes the form 
\begin{eqnarray}
&\qquad\hspace{-1cm} m_{eff}=4\pi{\int^R_0{\rho\,r^2\,dr}}=\frac{D\,AR^3\,(1-AR^2)^{n-2}}{2\,[1+D\,AR^2\,(1-AR^2)^{n-2}]}. \label{eq42}
\end{eqnarray}

Defining the compactification factor, $u(R)$ at $r=R$ as
\begin{eqnarray}
&\qquad\hspace{0cm} u(R)=\frac{m_{eff}(R)}{R}=\frac{D\,AR^2\,(1-AR^2)^{n-2}}{2\,[1+D\,AR^2\,(1-AR^2)^{n-2}]},   \label{eq43}
\end{eqnarray}
we find that the surface redshift $(Z_s)$ of the system is
\begin{eqnarray}
&\qquad\hspace{0cm} Z_s= \sqrt {{\frac {1-2\,nA{R}^{2}-A{R}^{2}}{1-A{R}^{2}}}}-1.  \label{eq44}
\end{eqnarray}

\begin{figure}[h!]
\centering
    \includegraphics[width=6.7cm]{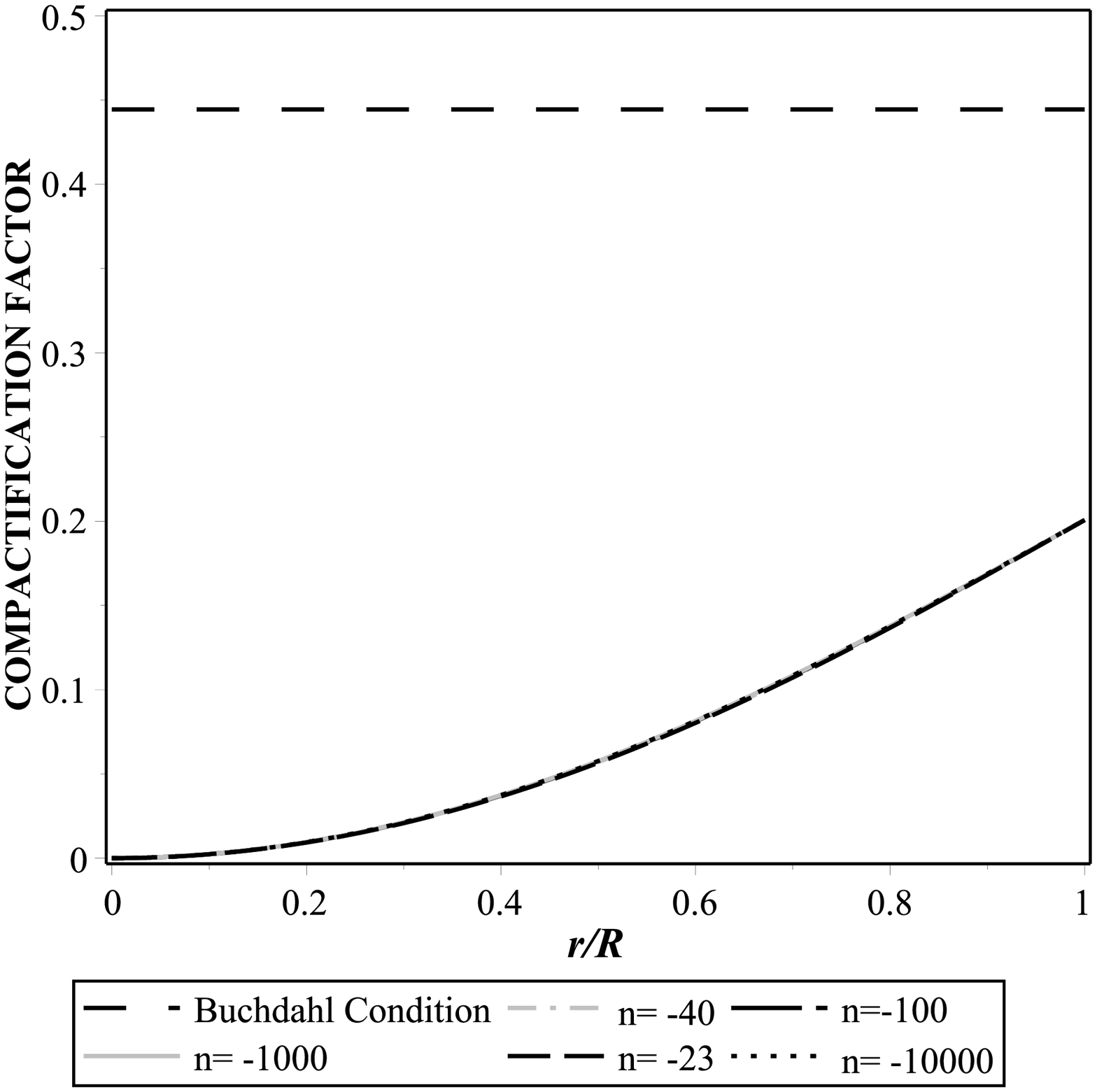}
    \includegraphics[width=6.7cm]{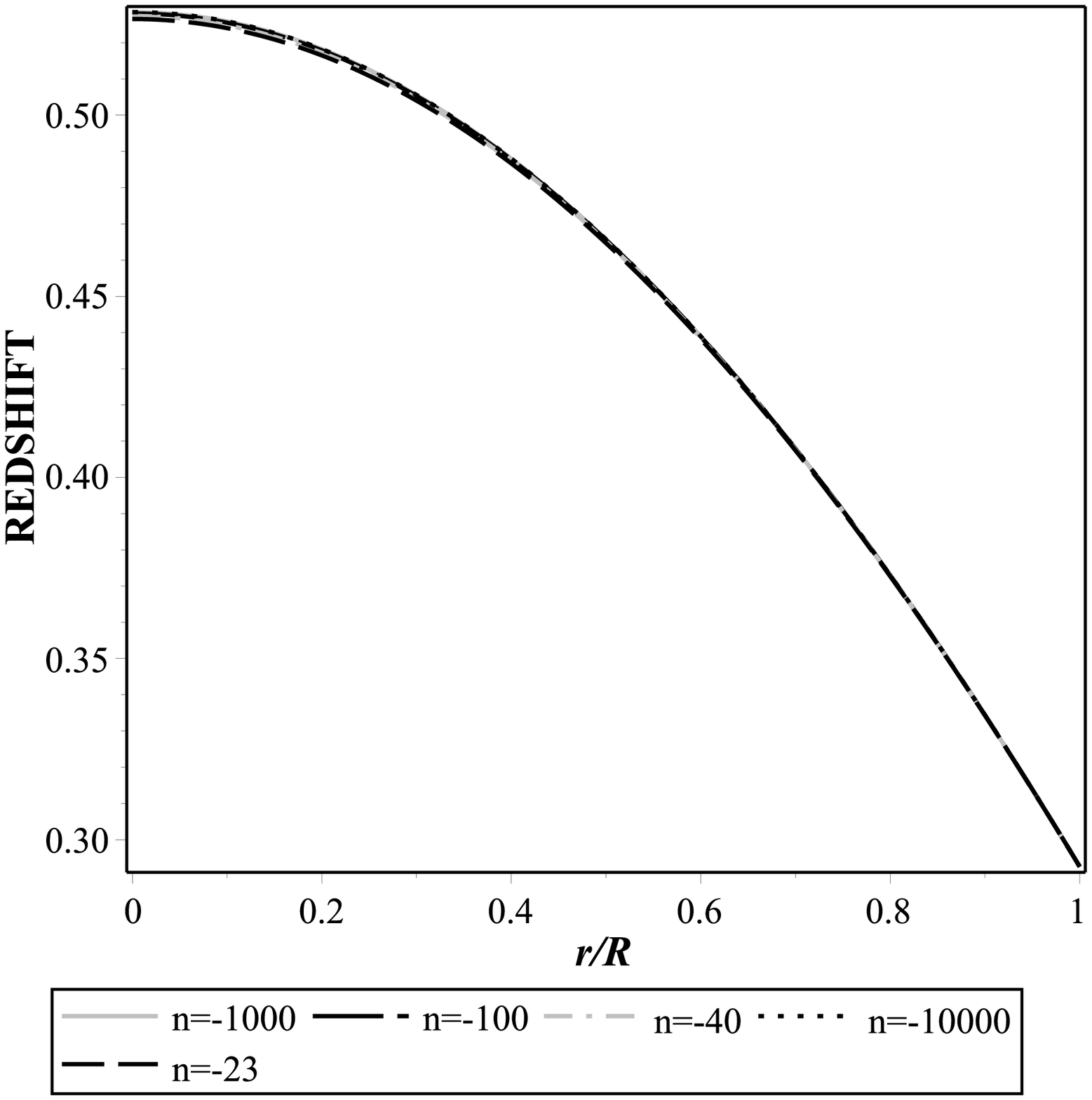}
\caption{Variation of the compactification factor ($u$) (left panel) and the redshift ($Z$) (right panel) with the fractional coordinate $r/R$ for $LMC\,X-4$. }
    \label{Fig11}
\end{figure}

The behavior of the compactification factor and redshift function with respect to the radial coordinate $r$ is shown in Fig.~\ref{Fig11}. One may predict from this figure that our model satisfies the Buchdahl condition~\cite{Buchdahl1959} well.

\section{Discussion and Conclusion}\label{sec7}
This paper discusses a generalized model for compact stars by starting with a general spherically symmetric metric of class
two and reducing it to a class one metric via a suitable choice of coordinates. Based on previous studies, it is assumed that
the highly dense spherically symmetric fluid sphere is anisotropic in nature.

After embedding the 4-dimensional spacetime in a $5d$ flat spacetime, the Einstein field equations are solved by
employing the physically relevant metric function $e^{\nu}=B\,(1-Ar^2)^n$, based on the analysis in Ref. \cite{Lake2003}.
The class one metric now leads to the determination of the metric function $e^{\lambda}$.  The computation of $\rho$,
$p_r$, and $p_t$ then yields the anisotropic factor $\Delta=p_t - p_r$, as well as the EOS parameters $\omega_r$ and
$\omega_t$.  This is followed by a detailed analysis of the boundary conditions that need to be satisfied by any
physically acceptable anisotropic solution.  In particular, for our anisotropic system the anisotropy is zero at the center and 
maximum at the surface as predicted by Deb et al.~\cite{Deb2016a}.  Values of various parameters were also determined from the boundary conditions. 

Other physical properties of the anisotropic solution are the regularity at the center and, equally important, the various
energy conditions, all of which were shown to be satisfied.

A thorough stability analysis was performed for all the compact models by means of the TOV equation, as well as the cracking
concept due to Herrera \cite{Herrera1992}, the latter depending on the square of the velocity of sound.  The subsequent
determination of the adiabatic index confirmed that the structures are stable.  A final topic is a discussion of the effective
gravitational mass in terms of the compactification factor and the surface redshift.  It was also shown that our model satisfies the Buchdahl condition.

One may find an interesting insight of our model that it is consistent with any compact stars like white dwarf or ultra dense compact stars. 
The generalized model is valid for any negative value of $n$. However, lower limit of $n$ for the white dwarf is $-3$ and for the ultra dense compact stars is $-23$.

Using the boundary conditions, the physical parameters needed for plotting were determined for various values of $n$ for the strange star $LMCX-4$ and summarized in Tables~1 and~2.  We find that high values of central density and central pressure of $LMCX-4$
 confirms that it is a probable candidate of strange star (see Table~2). Also we find the value of central density of the order $10^6~gm/{{cm}^3}$ and the central pressure of the order $10^{23}~dyne/{{cm}^2}$ for the white dwarf $Sirius~B$ (see Table~3) match well with the predicted result by Anchordoqui~\cite{Anchordoqui2016}. Thus our generalized model is appropriate to study any compact stellar configuration.


\begin{table}[t]
\centering \caption{Numerical values of physical parameters for the compact stars with the different values of $n$ } \label{Table 1}
\resizebox{\columnwidth}{!}{
\begin{tabular}{@{}cccccccc@{}} \hline

& & & $ n=-23$ & $n=-40$ & $n=-10^2$ & $n=-10^3$ &$n=-10^4$\\ \hline

Compact star & $M~(M_\odot)$ & $Radius$ & $A{R}^{2}$ & $A{R}^{2}$ & $A{R}^{2}$ & $A{R}^{2}$ & $A{R}^{2}$ \\ \hline
(Strange Star)& & & & & & & \\
 $LMC\,X-4$  &1.29$\pm$0.05~\cite{Rawls2011} & 9.48 ($km$)~\cite{Deb2016a}  & 14.369$\times 10^{-3}$  & 8.313$\times 10^{-3}$  &3.342$\times 10^{-3}$  &3.35$\times 10^{-4}$  & 3.35$\times 10^{-5}$ \\ \hline
& & & $ n=-3$ & $n=-10$ & $n=-10^2$ & $n=-10^3$ &$n=-10^4$\\ \hline
(White Dwarf)& & & & & & & \\
 $Sirius\, B$ & 1.034$\pm$0.026~\cite{HOLBERG1997} & 0.0084($R_{\odot}$)~\cite{HOLBERG1997} & 1.1353$\times10^{-5}$ & 6.528$\times10^{-6}$ & $2.6112\times10^{-6}$  & $2.6112\times10^{-7}$ & $2.6112\times10^{-8}$ \\ \hline 
\end{tabular}
}
\end{table}


\begin{table}[h!]
	\centering
	\caption{Physical Parameters for $LMC\,X-4$}\label{Table 2}
	\resizebox{\columnwidth}{!}{
	\begin{tabular}{cccccccccc} \hline
		value & Central Density & Surface Density & Central Pressure &  $A$ & $B$ & $K$ & $D$  \\
		of $n$ & $(gm/cm^{3}) $ & $(gm/cm^{3})$ & $(dyne/cm^{2})$ & $(km^{-2})$ & & $(km^2)$ &  \\ \hline
		$-23$  & 8.3530$\times 10^{14} $ & 6.3021$\times 10^{14} $ & 1.0386$\times 10^{35}$  & 1.598$\times 10^{-4}$  & 0.42911 & 895.5537 & 32.502   \\
		$-40$ & 8.4472$\times 10^{14} $ & 6.2671$\times 10^{14} $ & 1.0322$\times 10^{35}$ & 0.925$\times 10^{-4}$ & 0.42866 & 895.5154 & 56.813 \\
		$-10^2$ &  8.5253$\times 10^{14} $ & 6.2388$\times 10^{14} $ & 1.0270$\times 10^{35}$ & 3.719$\times 10^{-5}$ &  0.42826 & 895.4611 & 142.620 \\
		$-10^3$ & 8.5725$\times 10^{14} $ & 6.2213$\times 10^{14} $ & 1.0235$\times 10^{35}$ & 0.3730$\times 10^{-5}$ & 0.42806 & 895.4966 &  1429.807   \\
		$-10^4$ & 8.5765$\times 10^{14} $ & 6.2199$\times 10^{14} $ & 1.0231$\times 10^{35}$ & 0.03731$\times 10^{-5}$ & 0.42805 & 895.5236 & $14301.825$ \\ \hline

	\end{tabular}}
\end{table}

\begin{table}[h!]
	\centering
	\caption{Physical Parameters for $Sirius$ $B$}\label{Table 3}
		\resizebox{\columnwidth}{!}{
	\begin{tabular}{cccccccccc} \hline
		value & Central Density & Surface Density & Central Pressure &  $A$ & $B$ & $K$ & $D$  \\
		of $n$ & $(gm/cm^{3}) $ & $(gm/cm^{3})$ & $(dyne/cm^{2})$ & $(km^{-2})$ & & $(km^2)$ &  \\ \hline
		$-3$  & 2.4625$\times 10^{6}$ & 2.4622$\times 10^{6}$ &2.5729$\times 10^{23}$ & 2.5485$\times 10^{-12} $ & 0.99999249 & 2.6170$\times 10^{11} $ & 5.9979   \\
		$-23$ & 2.4628$\times 10^{6}$&  2.4619$\times 10^{6}$ &2.0139$\times 10^{23}$ &  3.3243$\times 10^{-13} $ &  0.99921709& 2.6171$\times 10^{11} $ & 45.9874 \\
		$-10^2$ &2.4626$\times 10^{6}$ & 2.4615$\times 10^{6}$ & 1.9490$\times 10^{23}$ &  7.6450$\times 10^{-14} $ & 0.99921708 &2.6174$\times 10^{11} $ & 199.9473  \\
		$-10^3$ &2.4626$\times 10^{6}$ & 2.4615$\times 10^{6}$ &1.9316$\times 10^{23}$ & 7.6450$\times 10^{-15} $ &0.99921708 &2.6175$\times 10^{11} $ & 1.9995 $\times 10^{3}$ \\
		$-10^4$ &2.4626$\times 10^{6}$ & 2.4615$\times 10^{6}$ &1.9299$\times 10^{23}$ & 7.6450$\times 10^{-16} $ &0.99921708 &2.6175$\times 10^{11} $ & 1.9995 $\times 10^{4}$  \\ \hline
	\end{tabular}}
\end{table}

\section*{Acknowledgments}
SKM acknowledges the support from the authorities of the University of
Nizwa, Nizwa, Sultanate of Oman. SR is thankful for the support from the Inter-University
Centre for Astronomy and Astrophysics (IUCAA), Pune, India, and the Institute of Mathematical Sciences (IMSc), 
Chennai, India for providing the working facilities and hospitality under the respective Associateship schemes.

\end{document}